\documentclass[twocolumn,amssymb, prl, aps]{revtex4-1}
\usepackage{graphicx}
\usepackage{amsmath}
\usepackage{braket}
\usepackage[colorlinks,linkcolor=black,citecolor=black,urlcolor=black,filecolor=black]{hyperref}
\begin{document}
\newcommand\numberthis{\addtocounter{equation}{1,2}\tag{\theequation}}
\title{Controlling many-body dynamics with driven quantum scars in Rydberg atom arrays}
\author{D. Bluvstein$^{1}$, A. Omran$^{1,2}$, H. Levine$^{1}$, A. Keesling$^{1}$, G. Semeghini$^{1}$, S. Ebadi$^{1}$, T. T. Wang$^{1}$, A.~A.~Michailidis$^{3}$, N. Maskara$^{1}$, W. W. Ho$^{1,4}$, S. Choi$^{5}$, M. Serbyn$^{3}$, M. Greiner$^{1}$, V. Vuleti\'{c}$^{6}$, M. D. Lukin$^{1}$}
\affiliation{$^1$Department of Physics, Harvard University, Cambridge, MA 02138, USA \\ $^2$ QuEra Computing Inc., Boston, MA 02135, USA \\ $^3$ IST Austria, Am Campus 1, 3400 Klosterneuburg, Austria \\ $^4$ Department of Physics, Stanford University, Stanford, CA 94305, USA \\ $^{5}$ Department of Physics, University of California Berkeley, Berkeley, CA 94720, USA \\ $^6$ Department of
Physics and Research Laboratory of Electronics,
Massachusetts Institute of Technology, Cambridge, MA 02139, USA} 

\begin{abstract}

Controlling non-equilibrium quantum dynamics in many-body systems is an outstanding challenge as interactions typically lead to thermalization and a chaotic spreading throughout Hilbert space. We experimentally investigate non-equilibrium dynamics following rapid quenches in a many-body system composed of 3 to 200 strongly interacting qubits in one and two spatial dimensions. Using a programmable quantum simulator based on Rydberg atom arrays, we probe coherent revivals corresponding to quantum many-body scars. Remarkably, we discover that scar revivals can be stabilized by periodic driving, which generates a robust subharmonic response akin to discrete time-crystalline order. We map Hilbert space dynamics, geometry dependence, phase diagrams, and system-size dependence of this emergent phenomenon, demonstrating novel ways to steer entanglement dynamics in many-body systems and enabling potential applications in quantum information science.

\end{abstract}

\maketitle

Dynamics of complex, strongly interacting many-body systems have broad implications in quantum science and engineering, ranging from understanding fundamental phenomena such as the nature of quantum gravity \cite{Maldacena2016} to realizing robust quantum information systems \cite{Arute2019,Zhong2020}. In these many-body systems, dynamics typically lead to a rapid growth of quantum entanglement and a chaotic spreading of the wave function throughout an exponentially large Hilbert space, a phenomenon associated with quantum thermalization \cite{Srednicki1994,Rigol2008,Kaufman2016}. Recent advances in the controlled manipulation of isolated, programmable many-body systems have enabled detailed studies of non-equilibrium states in strongly interacting quantum matter \cite{Schreiber2015,Langen2015,Kaufman2016}, in regimes inaccessible to numerical simulations on classical machines. Identifying non-trivial states for which dynamics can be stabilized or steered by external controls is a central question explored in these studies. 
For instance, it has been shown that strong disorder, leading to many-body localization (MBL), allows systems to suppress entanglement growth and retain memory of their initial state for long times \cite{Nandkishore2015}. Another striking example involves quantum many-body scars, which manifest as special initial states that avoid rapid thermalization within an otherwise chaotic system \cite{Heller1984,Bernien2017,Turner2018}. Further, periodic driving in strongly interacting systems can give rise to exotic non-equilibrium phases of matter, such as the discrete time crystal (DTC) which spontaneously breaks the discrete time-translation symmetry of the underlying drive \cite{Khemani2016,Else2016}.

\begin{figure}
\includegraphics[width=\columnwidth]{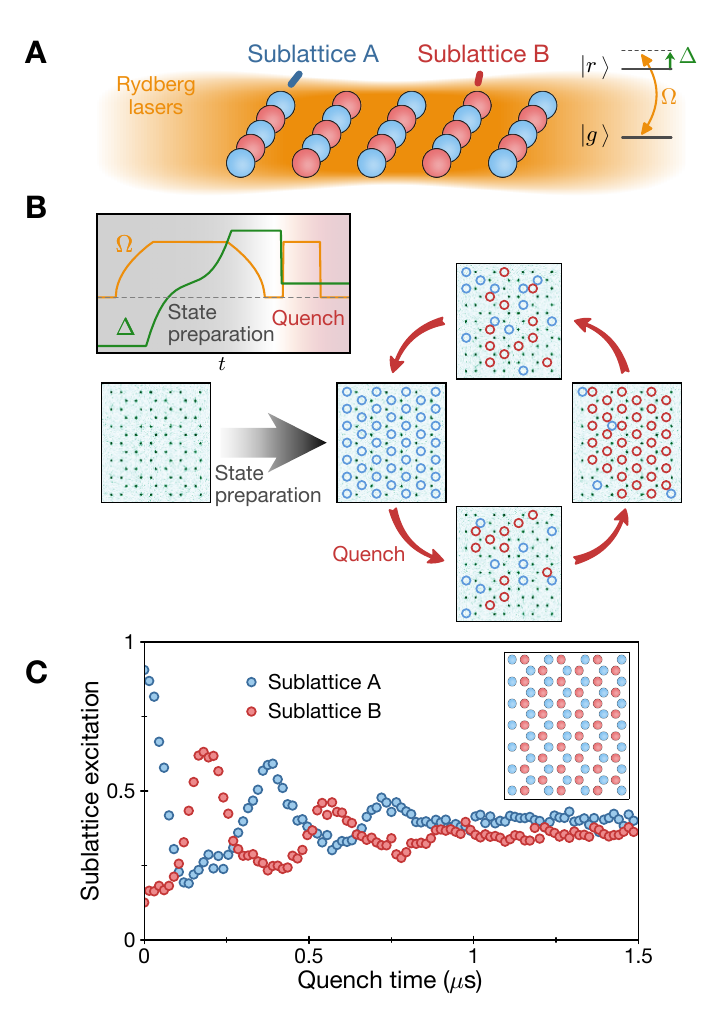}
\caption{\textbf{Experimental investigations of quantum many-body scars.} \textbf{(A)} Two-dimensional atom array subject to global Rydberg lasers with Rabi frequency $\Omega$ and detuning $\Delta$. \textbf{(B)} A quasi-adiabatic ramp of $\Delta$ and $\Omega$ prepares an antiferromagnetic state $\ket{\text{AF}_1}$ with sublattice A excited, and a detuning quench launches non-equilibrium dynamics. Atoms in $\ket{g}$ are imaged in optical tweezers via fluorescence while atoms in $\ket{r}$ (empty circles) are expelled and detected as atom loss. \textbf{(C)} The Rydberg population on sublattices $A$ and $B$ undergo periodic oscillations (Inset: geometry used here).
}
\label{fig1}
\end{figure}

In this Report, we investigate stability, thermalization, and control of 
quantum many-body scars in systems ranging from 3 to 200 strongly interacting qubits with varying geometry~\cite{Bernien2017,Ebadi2020}. We discover that entanglement dynamics associated with such scarring trajectories can be stabilized via parametric driving, resulting in an emergent phenomenon akin to discrete time-crystalline order. We show this phenomenon can be harnessed to steer entanglement dynamics in complex many-body systems.    

In our experiments, neutral $^{87}$Rb atoms are trapped in optical tweezers and arranged into arbitrary two-dimensional patterns generated by a spatial light modulator \cite{Labuhn2016,Ebadi2020}. This programmable system allows us to explore quantum dynamics in systems ranging from chains and square lattices to exotic decorated lattices, with sizes up to 200 atoms. All atoms are initialized in an electronic ground state $\ket{g}$ and coupled to a Rydberg state $\ket{r}$ by a two-photon optical transition with an effective Rabi frequency $\Omega(t)$ and detuning $\Delta(t)$, as depicted schematically in Fig.~1A. When excited into Rydberg states, atoms interact via a strong, repulsive van der Waals interaction $V \sim 1/d^6$, where $d$ is the inter-atomic separation, resulting in the many-body Hamiltonian \cite{Bernien2017},
\begin{align}
    \frac{H}{\hbar} = \frac{\Omega(t)}{2} \sum_i \sigma^x_i - \Delta(t) \sum_i n_i + \sum_{i<j} V_{ij} n_i n_j
\end{align}
where $\hbar$ is the reduced Planck constant, $n_i = \ket{r_i} \! \bra{r_i}$ is the projector onto the Rydberg state at site $i$ and $\sigma^x_i = \ket{g_i} \! \bra{r_i} +  \ket{r_i} \! \bra{g_i}$ flips the atomic state. We choose lattice spacings where the nearest-neighbor (NN) interaction $V_0 > \Omega$ results in the Rydberg blockade~\cite{Jaksch2000,Urban2009,Labuhn2016}, preventing adjacent atoms from simultaneously occupying $\ket{r}$. For large negative detunings, the many-body ground state is $\ket{g g g g ...}$, and at large positive detunings on bipartite lattices the ground state is antiferromagnetic, of the form $\ket{r g r g ...}$. Starting with all atoms in $\ket{g}$, adiabatically increasing $\Delta$ from large negative values to large positive values thus prepares antiferromagnetic initial states $\ket{\text{AF}}$ ~\cite{Pohl2010,Schauss2015,Bernien2017}; we choose array configurations (e.g. odd numbers of atoms) such that one of the two classical orderings, $\ket{\text{AF}_1}$, is energetically preferred.

To explore quantum scarring in two-dimensional systems, we prepare $\ket{\text{AF}_1}$ on an 85-atom honeycomb lattice, and then suddenly quench at fixed $\Omega$ to a small positive detuning (Fig.~1B). The system quickly evolves from $\ket{\text{AF}_1}$ into a disordered, vast superposition of many-body states as expected from a thermalizing system, but then strikingly the opposite order $\ket{\text{AF}_2}$ emerges at a later time \cite{Turner2018}. Through the same process the system evolves back to $\ket{\text{AF}_1}$, consistent with previous observations of quantum scars in one-dimensional chains~\cite{Bernien2017,Turner2018}. These scarring dynamics can be seen in the evolution of sublattice $A$ and $B$ populations as a function of quench duration (Fig.~1C), where disordered configurations arise when the sublattice populations are approximately equal. These observations are surprising in a strongly interacting system: the fact that the atoms entangle and disentangle periodically while traversing through the complicated Hilbert space (as shown theoretically \cite{Ho2019}) indicates a special dynamical behavior as well as a form of ergodicity breaking \cite{Turner2018,Ho2019}. This scarring behavior is captured by the so-called `PXP' model of perfect nearest-neighbor blockade, in which $V_0$ is infinite and interactions beyond nearest-neighbor are zero: $H_\text{PXP} = (\Omega/2) \sum_i P_{i-1} \sigma^x_i P_{i+1}$ with $P_i = \ket{g_i} \bra{g_i}$ \cite{Lesanovsky2012,Turner2018,Ho2019,Lin2019, Khemani2019}.

We observe this oscillatory behavior in a wide variety of bipartite lattices, shown in Fig.~2A (we do not observe scarring on the non-bipartite lattices we measure). As an example, we  plot the difference between the sublattice A and B populations $\braket{n}_A - \braket{n}_B$ for a 49-atom square and a 54-atom decorated honeycomb \cite{Michailidis2020b}, with Rabi frequency $\Omega/2\pi = 4.2$ MHz and interaction strength $V_0/2\pi = 9.1$ MHz. We note a marked difference in the lifetime of periodic revivals for these two different lattices. Quantitatively, we find that dynamics of $\braket{n}_A - \braket{n}_B$ are well-described by a damped cosine, $y_0 + C \cos(\tilde{\Omega} t) \exp(-t/\tau)$, with oscillation frequency $\tilde{\Omega}$, decay time $\tau$, offset $y_0$, and contrast $C$. While $\tilde{\Omega} \approx 0.6~ \Omega$ on both the square and decorated honeycomb lattices, the fitted $\tau$ for these two different configurations are 0.22(1) $\mu$s and 0.50(1) $\mu$s, respectively.

To understand this geometry dependence, we consider an empirical model for the decay rate of many-body scars (see \cite{Supplement}), parametrized as follows:
\begin{align}
\frac{1}{\tau} = \alpha \left(\frac{1}{2\pi} \sum_{\text{NN}} \frac{\Omega^2}{4 V_0}\right) + \beta \left(\frac{1}{2\pi} \sum_{\text{NNN}} V_{ij} \right) + \frac{1}{\tau_0}
\label{decay}
\end{align}
where the first two terms capture deviations of the
Rydberg Hamiltonian from the idealized PXP model, due to second-order virtual coupling to states violating blockade and next-nearest-neighbor (NNN) interactions, respectively \cite{Supplement}; $\alpha, \beta, \tau_0$ are phenomenological values. In Fig.~2B we plot the measured $1/\tau$ as a function of the first and second terms in Eq.~\ref{decay} for all geometries shown in Fig.~2A and varied interaction strengths $V_0$. We find that the decay rates fit well to a plane with slopes $\alpha = 0.72(12)$ and $\beta = 0.58(5)$ and offset $1/\tau_0 = 0.4(2)$ MHz. Note that $1/\tau_0 \ll 1/\tau$, i.e., we find that the decay of scars is dominated by imperfect blockade and long-range interactions. The observation that long-range fields contribute to decay also motivates quenching to small positive $\Delta_q = \Delta_{q,\text{opt}} = 1/2 \sum_{i,j > \text{NN}} V_{ij}$, which enhances scarring by cancelling the static, mean-field contribution from the long-range interactions \cite{Supplement}, and is implemented for all geometries throughout this work. These results also suggest an intrinsic limit to the scar lifetime, coming from the trade-off between imperfect blockade ($\propto 1/V_0$) and long-range interactions ($\propto V_0$). E.g., with $\Omega/2\pi = 4.2$~MHz, for a one-dimensional chain at an optimal $V_0/2\pi \approx 19$ MHz we estimate a maximum lifetime $\tau_{\text{max}} \approx 0.9\, \mu$s, or instead $\tau_{\text{max}} \approx 0.4 \, \mu$s for a honeycomb lattice.

\begin{figure}
\includegraphics[width=\columnwidth]{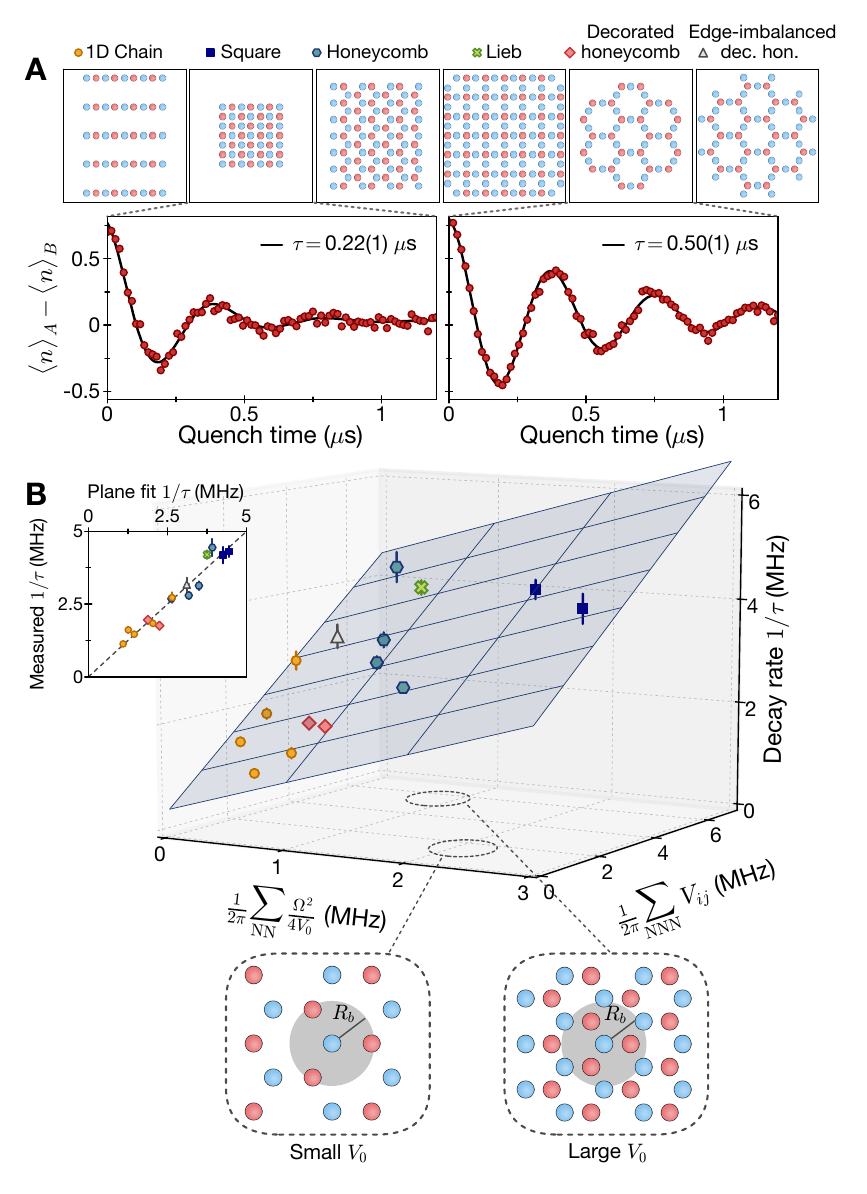}
\caption{\textbf{Universal empirical description of scar lifetime.} \textbf{(A)} Different geometries used in this study. The lifetime $\tau$ of the sublattice excitation difference depends strongly on the geometry. \textbf{(B)} As a function of coupling to blockade-violating states ($\propto\Omega^2/V_0$) and  next-nearest-neighbor (NNN) interactions, the scar decay rate $1/\tau$ displays a bilinear dependence (Inset: cross-section of the plane). Schematics depict regimes where the two different decay processes dominate.
}
\label{fig2}
\end{figure}

\begin{figure*}
\includegraphics[width=2\columnwidth]{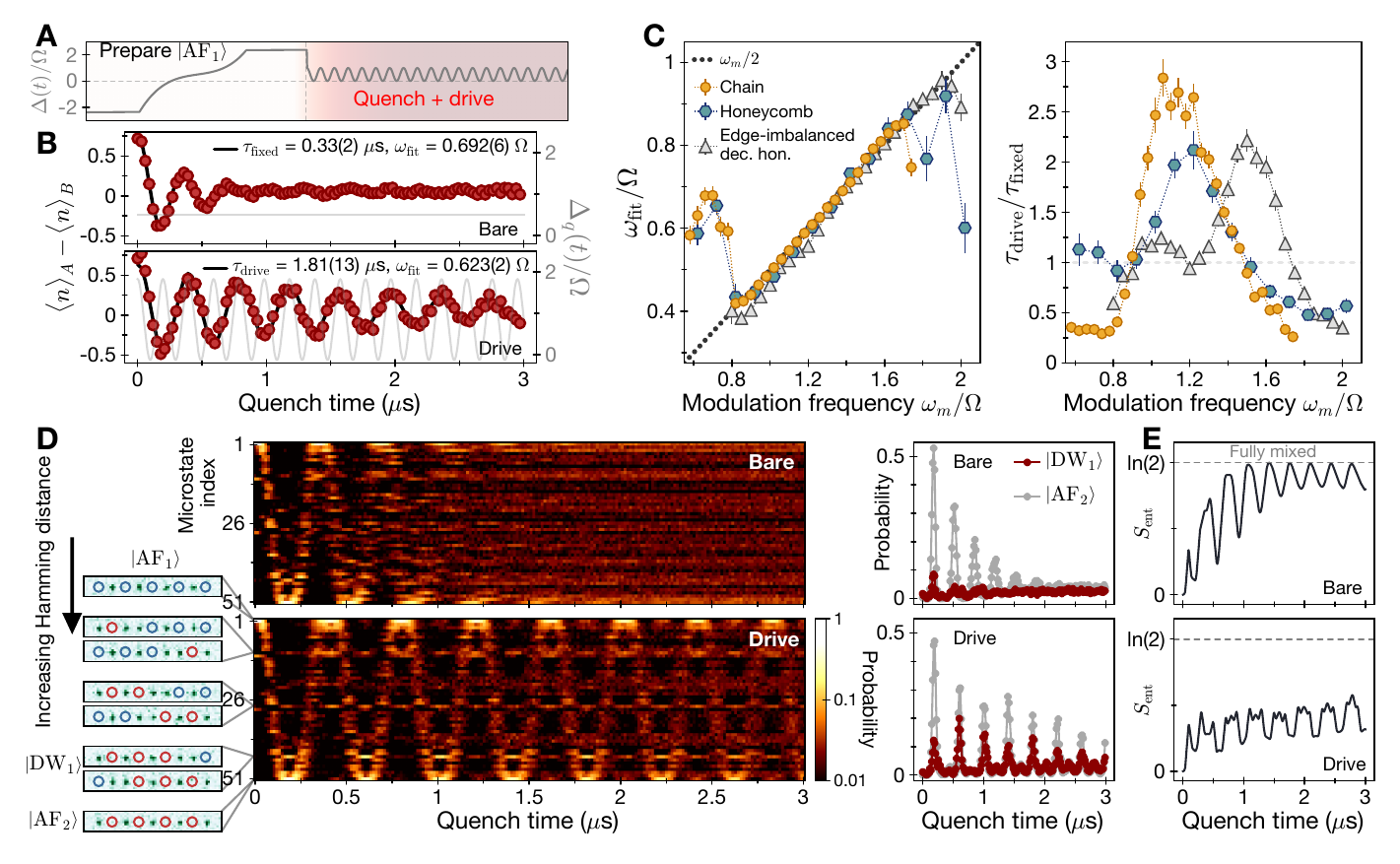}
\caption{\textbf{Emergent subharmonic locking and stabilization.} \textbf{(A)} Pulse sequence showing state preparation and quench with $\Delta_q(t)$. \textbf{(B)} Scar dynamics on a chain during quench to fixed optimal detuning (bare) with lifetime $\tau_{\text{fixed}}$, and time-dependent detuning (drive) with modulation frequency $\omega_m$ = 1.24 $\Omega$ and lifetime $\tau_{\text{drive}}$. The drive increases the scar lifetime and changes its frequency to $\omega_m/2$. \textbf{(C)} Scar lifetime and response frequency as a function of $\omega_m$, showing a lifetime increase and subharmonic locking. \textbf{(D)} Dynamics of the entire Hilbert space measured with experimental snapshots (0.5 million total bit strings). The microstates of the constrained Hilbert space are ordered by $n_A - n_B$, or equivalently by Hamming distance from $\ket{\text{AF}_1}$ \cite{Supplement}. Right subplots highlight $\ket{\text{AF}_2}$ and a state with a domain wall $\ket{\text{DW}_1}$. \textbf{(E)}~Reduced density matrix of a single atom in a chain (numerics) shows that driving reduces the growth of entanglement entropy $S_{\text{ent}}$.
}
\label{fig3}
\end{figure*}

We next investigate the effect of parametric driving on many-body scars. To this end, we implement quenches to a time-dependent detuning $\Delta_q(t) = \Delta_0 + \Delta_m \cos(\omega_m t)$, as illustrated in Figure 3A, and explore a non-perturbative regime of $\Delta_m, \Delta_0, \omega_m \sim \Omega$. Remarkably, in Fig.~3B we find that such a quench results in a five-fold increase of scar lifetime compared to the fixed-detuning case, for properly chosen drive parameters (modulation frequency $\omega_m = 1.24 ~\Omega$, offset $\Delta_0 = 0.85~ \Omega$, and amplitude $\Delta_m = 0.98 ~\Omega$ for this 9-atom chain). Further, we find the drive changes the oscillation frequency $\tilde{\Omega}$ to $\omega_m/2$, apparent in the synchronous revival of $\braket{n}_A - \braket{n}_B$ every two drive periods of $\Delta_q(t)$.

Figure~3C shows the scar lifetime and oscillation frequency as a function of modulation frequency $\omega_m$, for a 9-atom chain (with different $V_0$ than Fig.~3A), a 41-atom honeycomb, and a 66-atom edge-imbalanced decorated honeycomb (tabulation of system and drive parameters in \cite{Supplement}). For all three lattices, a robust subharmonic locking of the scar frequency is observed at $\omega_m/2$ over a wide range of $\omega_m$, accompanied by a marked increase in the scar lifetime. We note that significant lifetime enhancements are found even when $\Delta_m, \Delta_0 \gg \sum_{\text{NNN}} V_{ij}$, and even in numerics for the idealized PXP model \cite{Supplement}, indicating that the physical origin of the enhancement is not simply a mean-field-interaction cancellation akin to fixed~$\Delta_{q,\text{opt}}$.

To gain insight into the origin of the subharmonic stabilization, Figure~3D shows the experimentally observed distribution of microscopic many-body states across the entire Hilbert space of the 9-atom chain, as a function of quench time. 
For the fixed detuning quench, oscillations between $\ket{\text{AF}_1}$ and $\ket{\text{AF}_2}$ product states are observed, before the quantum state spreads and thermalizes to a near-uniform distribution across the many-body states \cite{Srednicki1994,Rigol2008}. Notably, parametric driving not only delays thermalization, but also alters the actual trajectory being stabilized: the driven case also shows periodic, synchronous occupation of several other many-body states, seemingly dominated by those with near-maximal excitation number (indicated in the left panel of Fig.~3D). This suggests that, rather than enhancing oscillations between the $\ket{\text{AF}}$ states, the parametric driving actually stabilizes the scar dynamics to oscillations between entangled superpositions composed of various product states. Figure~3E further illustrates the change in trajectory with numerical simulations of the local entanglement entropy, revealing that driving stabilizes the periodic entangling and disentangling of an atom with the rest of the system.

We observe this emergent subharmonic stabilization for a wide range of system and drive parameters. Figs.~4A and 4B show the time dynamics of $\braket{n}_A - \braket{n}_B$ and the normalized intensity of its associated Fourier transform $\left|S(\omega)\right|^2$ as a function of the drive frequency for a 9-atom chain. A response is observed at $\omega = \omega_m$ for $\omega_m < 0.8 ~\Omega$, before suddenly transitioning into a subharmonic response $\omega = \omega_m / 2$ for $\omega_m > 0.8 ~\Omega$. For different drive parameters a weak $4^{\text{th}}$ subharmonic response at $\omega = \omega_m / 4$ is also observed \cite{Supplement}. To quantify the robustness of the observed response, we evaluate the subharmonic weight, $|S(\omega=\omega_m/2)|^2$, which encapsulates both the $\omega_m/2$ response and enhanced lifetime \cite{Zhang2017,Choi2017}. Fig.~4C shows the corresponding results for a 9-atom chain and a 41-atom honeycomb as a function of the modulation frequency $\omega_m$ and the lattice spacing $a$ (in units of the blockade radius $R_b$ defined by $V(R_b) = \Omega$). A wide plateau in the subharmonic weight is clearly observed for both lattices, as a function of both modulation frequency and interaction strength (range $0.6 - 0.9~a/R_b$ corresponds to $V_0/2\pi \approx 8 - 80$ MHz). To quantify the many-body nature of this stable region \cite{Else2016}, we define the subharmonic rigidity, which evaluates the robustness of the subharmonic response over a range of modulation frequencies and is defined as $\sum_{\omega_m} |S_{\omega_m}(\omega=\omega_m/2)|^2$ for $\omega_m = 0.75, 0.85, ..., 1.75~ \Omega$. Figure~4D plots subharmonic rigidity vs system size for both a chain and a honeycomb lattice, increasing with system size until saturating at roughly 13 atoms, and appearing stable for the honeycomb lattice even to 200 atoms.

We now turn to a discussion of these experimental observations. 
The emergent subharmonic response and its rigidity is strongly reminiscent of those associated with discrete time-crystalline order~\cite{Khemani2016,Else2016,Zhang2017,Choi2017,Yao2020}. Yet, there are clear distinctions. Specifically, this behavior is observed only for antiferromagnetic initial states, while other initial states such as $\ket{ggg ... }$ thermalize and do not show subharmonic responses \cite{Supplement}. This significant state dependence distinguishes these observations from conventional MBL or prethermal time crystals \cite{Else2017}, where subharmonic responses are not tied to special initial states. Moreover, it is striking that our drive, whose frequency is resonant with local energy scales, enhances quantum scarring and ergodicity breaking instead of rapidly injecting energy into the system, as would generally be expected in many-body systems \cite{Ponte2015}.

\begin{figure*}
\includegraphics[width=104mm]{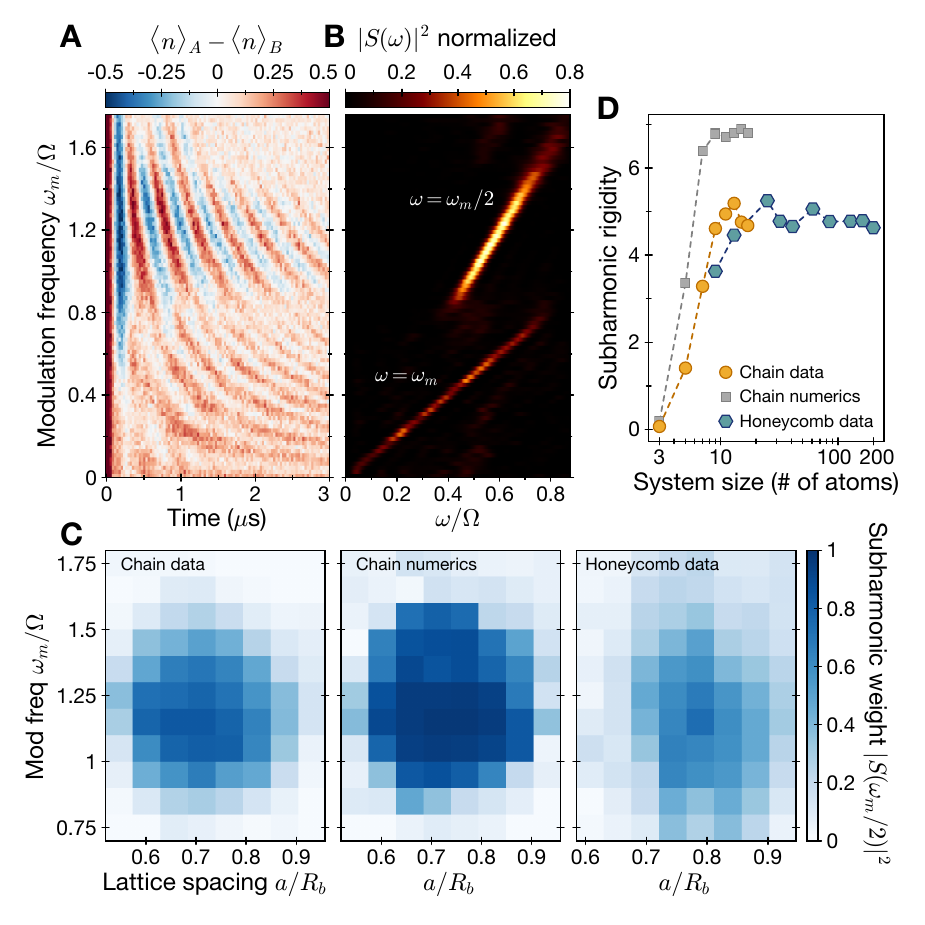}
\caption{\textbf{Robustness of the subharmonic response.} \textbf{(A}) Dynamics of sublattice population difference after quench, as a function of modulation frequency. \textbf{(B)} Fourier transform intensity $|S(\omega)|^2$ of data in (a), showing a harmonic locking for $\omega_m < 0.8 ~\Omega$ and a subharmonic locking for $\omega_m > ~ 0.8 ~\Omega$. \textbf{(C)} Phase diagram of the subharmonic response $|S(\omega=\omega_m/2)|^2$ in chain data (left), chain numerics (middle) from perfectly initialized $\ket{\text{AF}_1}$ without experimental imperfections, and honeycomb data (right). \textbf{(D)} Increase of subharmonic rigidity (see text) with increasing system size.
}
\label{fig4}
\end{figure*}

To gain intuition into the origin of our experimental observations, we consider a toy, pulsed driving model with Floquet unitary $U_F(\theta, \tau) = e^{-i\theta \sum_i n_i}e^{-i H_{\text{PXP}} \tau}$, where $\theta$ arises from an infinitesimal, strong detuning pulse. Due to the particle-hole symmetry of the PXP Hamiltonian, for $\theta=\pi$ the time evolution $e^{-i H_{\text{PXP}} \tau}$ during one pulse is cancelled by the time evolution $e^{i H_{\text{PXP}} \tau}$ in a subsequent pulse, generating an effective many-body echo and subharmonic response \cite{Supplement}. Interestingly, for small deviations from perfect $\pi$ rotations,  $\theta = \pi + \varepsilon$, revivals vanish for generic initial states but persist robustly for an initial $\ket{\text{AF}}$ state \cite{Supplement}. This behavior can be understood as follows. Due to the scarring character of the antiferromagnetic initial states, the PXP evolution approximately realizes an effective $\pi$-pulse from $\ket{\text{AF}_1}$ to $\ket{\text{AF}_2}$, but results in ergodic spreading for other initial states. Accordingly, for $\theta = \pi + \varepsilon$, evolution still approximates a many-body echo for the scarred $\ket{\text{AF}}$ but does not reverse the chaotic evolution of generic initial states. Finally, the additional $\varepsilon \sum_i n_i$ in fact serves as a ``stabilizing Hamiltonian'' by creating an effective gap between the $\ket{\text{AF}}$ states (which have maximal atomic excitations $n_\text{max} = \sum_i n_i$) from the rest of the spectrum. In practice, the $\ket{\text{AF}}$ states will be dressed by other states with near-maximal atomic excitations, consistent with Fig.~3D showing stabilized oscillations between two superpositions of states with largest $\sum_i n_i$. Although the above arguments utilize pulses, neglect large NNN interactions, and do not explicitly explain the observations in imbalanced lattices (Fig.~3C), this analysis already offers useful insight and warrants further study.

These considerations indicate that the observed subharmonic stabilization of many-body scars in large-scale quantum systems constitutes a new physical phenomenon that can be used for steering quantum entanglement dynamics in complex systems. While these observations challenge conventional understandings of quantum thermalization, the exact nature and conditions for these phenomena and their relationship to dynamical phases of matter such as the DTC warrant further theoretical and experimental investigation. In particular, it would be interesting to explore if many-body states with larger degrees of entanglement could also be stabilized by driving. Such studies could be extended to systems with more complex geometry, control, and topology: ranging from other initial states \cite{Mukherjee2020b}, non-bipartite arrays \cite{Labuhn2016}, and utilizing hyperfine qubits \cite{Levine2019}, to implementing these techniques in other controllable many-body systems. This phenomenon opens the door to tantalizing possibilities for robust creation and control of complex entangled states in the exponentially large Hilbert spaces of many-body systems, with intriguing potential applications in areas such as quantum metrology \cite{Giovannetti2004} and quantum information science \cite{Maldacena2016,Arute2019,Zhong2020,Monz2011}.

\textbf{Acknowledgements}
\\
We thank many members of the Harvard AMO community, particularly Elana Urbach, Samantha Dakoulas, and John Doyle for their efforts enabling safe and productive operation of our laboratories during 2020. We thank D. Abanin, I. Cong, F. Machado, H. Pichler, N. Yao, B. Ye, and H. Zhou for stimulating discussions. \textbf{Funding:} We acknowledge financial support from the Center for Ultracold Atoms, the National Science Foundation, the Vannevar Bush Faculty Fellowship, the U.S. Department of Energy, the Office of Naval Research, the Army Research Office MURI, and the DARPA ONISQ program. D.B. acknowledges support from the NSF Graduate Research Fellowship Program (grant DGE1745303) and The Fannie and John Hertz Foundation. H.L. acknowledges support from the National Defense Science and Engineering Graduate (NDSEG) fellowship. G.S. acknowledges support from a fellowship from the Max Planck/Harvard Research Center for Quantum Optics. T.T.W. acknowledges support from Gordon College. A.M. and M.S. were supported by European Research Council (ERC) under the European Union’s Horizon 2020 research and innovation program (Grant Agreement No.~850899). N.M. acknowledges support by the Department of Energy Computational Science Graduate Fellowship under Award Number(s) DE-SC0021110. W.W.H. is supported by the Moore Foundation’s EPiQS Initiative Grant No.~GBMF4306, the NUS Development Grant AY2019/2020, and the Stanford Institute of Theoretical Physics. S.C. acknowledges support from the Miller Institute for Basic Research in Science.

\bibliographystyle{Science}
\bibliography{library.bib}

\clearpage
\onecolumngrid
\begin{center}
    \textbf{\Large Supplementary Materials}
\end{center}
\normalsize

\setcounter{equation}{0}
\setcounter{figure}{0}
\setcounter{table}{0}
\makeatletter
\renewcommand{\theequation}{S\arabic{equation}}
\renewcommand{\thefigure}{S\arabic{figure}}
\renewcommand{\thetable}{S\arabic{table}}

\tableofcontents


\section{1. Experimental setup and details \label{Sec:expt}}

We initialize a sorted array of atoms in a desired geometry and optically pump the atoms into the stretched state $\ket{5S_{1/2}, F=2, m_F=-2}$. The atoms are then illuminated by two Rydberg laser beams at $1013\,$nm and $420\,$nm, with single-photon Rabi frequencies of $\Omega_{1013}/(2\pi)\approx 50\,$MHz and $\Omega_{420}/(2\pi)\approx 160\,$MHz and a detuning from the $6P_{3/2}$ intermediate state of $\delta/(2\pi)\approx 1\,$GHz. Using an arbitrary waveform generator (AWG) connected to an acousto-optic modulator (AOM), we control the intensity, frequency, and phase of the 420-nm light arbitrarily. We apply the 420-nm light such that the two-photon detuning $\Delta$ starts at a large negative value, and sweep to large positive values using a cubic time profile. For each geometry, we optimize the sweep parameters to maximize the state preparation fidelity, as measured by the contrast between Rydberg populations on sublattices $A$ and $B$. See \cite{Ebadi2020} for a detailed, up-to-date characterization of our experimental apparatus and adiabatic state preparation in two-dimensional arrays.

\section{2. Thermalization mechanisms and fixed-detuning quenches \label{Sec:therm}}

\subsection{2.1. Derivation of effective Hamiltonian\label{sec:SW}}

The Rydberg blockade mechanism arises in the limit of strong nearest-neighbor interactions, $V_0\gg\Omega$, such that the many-body Hilbert space is split into disconnected sectors distinguished by the total number of nearest-neighbor excitations \cite{Abanin2017a}. In this section we employ Schrieffer-Wolff (SW) perturbation theory to derive an effective Hamiltonian in the sector of zero nearest-neighbor excitations starting from the Rydberg Hamiltonian, defined in  Eq.~(1) in the main text. The effective Hamiltonian is obtained from an expansion in the small parameter $\Omega/V_0$ up to second order. We describe the main steps of the expansion, applicable in any lattice geometry.
The subleading terms in the effective Hamiltonian provide important insights into the physical processes that facilitate thermalization of the system at short timescales and will be used in Section~\ref{Sec:DecayRate} to justify the expression for the empirical decay rate of scars defined in Eq.~(2) of the main text.

The first step of the SW transformation consists of the splitting of the full Hamiltonian into the dominant part ($H_{0}$) and the perturbation ($Q$) so that $H = H_0 + Q$. We consider the limit where the nearest-neighbor interaction strength $V_0$ is the dominant energy scale compared to Rabi frequency $\Omega$, detuning $\Delta$, and longer-range interactions. This naturally leads to the following splitting:
\begin{equation}
H_{0} = V_0 
\sum_{\langle ij \rangle } n_{i}n_{j},\qquad Q = \frac{\Omega}{2}\sum_{i}\sigma^{x}_{i} - \Delta\sum_{i}n_{i} + \frac{V_{0}}{2}\sum_{i,j > \text{NN}}\frac{n_{i}n_{j}}{(d_{ij}/a)^6},
\end{equation}
where $d_{ij}/a$ is the distance between sites $i$ and $j$ normalized by the nearest-neighbor spacing $a$, and the last term sums over all sites $i,j$ with $d_{ij}/a > 1$ (i.e. beyond nearest neighbors), with the factor of $1/2$ accounting for double-counting of pairs.

The unperturbed Hamiltonian $H_0$ effectively counts the total number of nearest-neighbor excitations in the system. We further split the perturbation $Q$ into the sum of generalized ladder operators $T_{m}$, defined so that $[H_{0},T_{m}] = m V_{0} T_{m}$, with $m$ being an integer. Physically, this commutation rule implies that the operator $T_{m}$ increases energy by  $m V_{0}$ when applied to an eigenstate of $H_0$. For the Rydberg Hamiltonian, the integer $m$ identifies the number of nearest-neighbor excitations that are either created, if $m> 0$, or annihilated, if $m<0$, by the application of $T_{m}$ to an eigenstate of $H_{0}$. The detuning as well as the longer range interactions commute with the dominant term in the Hamiltonian $H_0$ and therefore, contribute only to the $T_{0}$ operator,
\begin{equation}
T_{0} = \frac{\Omega}{2}\sum_{i}\mathcal{P}_{i}^{D,0}\sigma^{x}_{i} - \Delta\sum_{i}n_{i}+\frac{V_{0}}{2}\sum_{i,j > \text{NN}}\frac{n_{i}n_{j}}{(d_{ij}/a)^6}.
\end{equation}
The remaining ladder operators $T_{m\neq 0}$ originate from the action of the $(\Omega/2)\sigma^x$ term,
\begin{equation}
T_{m} = \frac{\Omega}{2}\sum_{i}\mathcal{P}_{i}^{D,m}\sigma^{+}_{i} \quad \text{for}\quad m=1,\ldots,D\quad \text{with} \qquad T_{-m}= T^\dag_{m},
\end{equation}
where $D$ is the number of nearest neighbors for the given lattice and  calligraphic operators $\mathcal{P}^{D,m}_{i}$ are defined as projectors onto the subspace where $m$ nearest neighbors of site $i$ are simultaneously excited. If the Rydberg atom at site $i$ is flipped in this subspace, the energy of the state measured with respect to $H_0$ will change proportionally to the number of excited nearest neighbors $m$, as desired.

The SW transformation of order $l$ is a rotation of the Hamiltonian, $H^{(l)}=\mathcal{U}^{\dag}_{l}H\mathcal{U}_{l}$ that eliminates all off-diagonal (in the unperturbed eigenbasis) operators up to $O(\Omega^{l+1}/V_{0}^l)$. The generator of the SW transformation at order $l=1$ can be written as $\mathcal{U}_{1} = \text{exp}(-\sum_{m\neq 0}\frac{T_{m}}{m V_{0}})$. Higher-order generators have a more complicated form, containing nested commutators of the generalized ladder operators. The rotated Hamiltonians $H^{(l)}$ are truncated at $O(\Omega^{l+1}/V_{0}^l)$ and therefore, the equalities below are defined up to the truncation order. The first-order Hamiltonian is,
\begin{equation}
H^{(1)} = H_{0} + T_{0}  =
V_0 \sum_{\langle ij\rangle} n_i n_j
+  \frac{\Omega}{2}\sum_{i}\mathcal{P}_{i}^{D,0}\sigma^{x}_{i} - \Delta\sum_{i}n_{i}+\frac{V_{0}}{2}\sum_{i,j > \text{NN}}\frac{n_{i}n_{j}}{(d_{ij}/a)^6}.
\label{eqn:first_order_H}
\end{equation}
The first term $H_0 = V_0 \sum_{\langle ij\rangle} n_i n_j$ contributes a constant that is equal to zero, as we restrict to the so-called `Rydberg-blockaded' Hilbert space in which no two neighboring sites are simultaneously excited. The Hamiltonian~(\ref{eqn:first_order_H}) is an effective Hamiltonian in the Rydberg-blockaded Hilbert space. In particular, the projector $\mathcal{P}^{D,0}_{i,j}$ that dresses the spin-flip operator $\sigma^x$ ensures that Rydberg excitations obey the blockade condition, leading to the presence of a kinetic constraint in the dynamics. Equation~\eqref{eqn:first_order_H} is equivalent to the ``PXP-model''~\cite{Lesanovsky2012,Turner2018} but in the presence of detuning and long-range interactions.

To probe additional thermalization processes that stem from virtual excitations that violate Rydberg blockade, we consider the effective Hamiltonian with terms up to second order,
\begin{equation}\label{eqn:H2fin}
H^{(2)} = H^{(1)} + \sum^{D}_{m=1}\frac{[T_{m},T_{-m}]}{m V_{0}} = H^{(1)} + \frac{\Omega^2}{4V_{0}}\left(\sum_{i}\sum^{D}_{m=1}\frac{1}{m}\mathcal{P}_{i}^{D,m}\sigma^{z}_{i} - \sum_{ \langle ij\rangle } 
\mathcal{P}^{D,0}_{i,j}(\sigma^{+}_{i}\sigma^{-}_{j} + \text{H.c.}) \right).
\end{equation}
Where the first term in parenthesis corresponds to multi-site interactions and the second term describes kinetically constrained hopping of Rydberg excitations between nearest-neighbor sites $i,j$ provided that all neighbors of these two sites are in the $\ket{g}$ state.

Collecting all terms  together we obtain the final expression for the effective Hamiltonian:
\begin{equation}
H^{(2)} = \frac{\Omega}{2}\sum_{i}\mathcal{P}_{i}^{D,0}\sigma^{x}_{i} - \Delta\sum_{i}n_{i}+\frac{V_{0}}{2}\sum_{i,j > \text{NN}}\frac{n_{i}n_{j}}{(d_{ij}/a)^6} + \frac{\Omega^2}{4V_{0}}\left(\sum_{i}\sum^{D}_{m=1}\frac{1}{m}\mathcal{P}_{i}^{D,m}\sigma^{z}_{i} - \sum_{ \langle ij\rangle } 
\mathcal{P}^{D,0}_{i,j}(\sigma^{+}_{i}\sigma^{-}_{j} + \text{H.c.}) \right).
\label{eqn:SW}
\end{equation}
Previous theoretical studies have predominantly focused on the 
long-lived oscillations from $\ket{\text{AF}}$-type initial states in the pure PXP-model that is given by the first term in $H^{(2)}$. The presence of quantum many-body scars in this Hamiltonian, discussed in one-dimensional chains~\cite{Turner2018} and generic bipartite two-dimensional lattices~\cite{Michailidis2020,Michailidis2020b}, leads to long intrinsic decay timescales of the oscillations of local observables. It is thus reasonable to assume that the decay  rates seen in experiments (and numerics of the full Rydberg Hamiltonian) are caused by the remaining terms in Eq.~(\ref{eqn:H2fin}) that describe deviations from the PXP  model, as such deformations are observed to generally increase thermalization rates~\cite{Turner2018,Khemani2019}. The derivation of the second-order Hamiltonian $H^{(2)}$ for the Rydberg-blockaded Hilbert space demonstrates that the following microscopic mechanisms dominate deviations from the PXP-model: (i) detuning that is controlled experimentally by the parameter $\Delta$, (ii) longer-range interactions that have overall magnitude scaling with $V_0$, but strongly depend on the geometry of the lattice, and (iii) higher-order corrections that scale as $\Omega^2/4V_0$. These terms will be used in Section~\ref{Sec:DecayRate} to justify the phenomenological model for thermalization rate used in the main text (see also Eq.~\eqref{eqn:appendix_thermalization}). 

\subsection{2.2. Optimal fixed global detuning for suppressing long-range interactions\label{sec:detune}}

In this section we show that there is an astute choice of detuning $\Delta_q$ such that the detrimental effect of long-range interaction terms is partially mitigated. As discussed in the previous section and in the main text, we find empirically for fixed-detuning quenches that deviations from the pure PXP Hamiltonian limit the lifetime of the scars we observe. This motivates the rationale for quenching to small positive values of $\Delta_q$ as opposed to $\Delta_q = 0$, as the long-range interactions are always positive and so can be partially compensated by a fixed detuning. Mathematically, the optimal value of detuning can be deduced from rewriting the second and third terms in Eq.~(\ref{eqn:SW}) via the spin operator $S^z_i = (1/2) \sigma^z_i$ such that $n_i = S^z_i + 1/2$, giving
\begin{equation}
- \Delta\sum_{i}n_{i}+\frac{V_{0}}{2}\sum_{i,j > \text{NN}}\frac{n_{i}n_{j}}{(d_{ij}/a)^6}
=
  \frac{1}{2} \sum_{i,j > \text{NN}} V_{ij} S^z_i S^z_j + \sum_i S^z_i \left(-\Delta + \frac1{2}\sum\limits_{\substack{i,j > \text{NN}}} V_{ij}\right),
\end{equation}
where $d_{ij}/a$ is the distance between sites $i$ and $j$ normalized by the nearest-neighbor spacing $a$. We observe that terms proportional to $S^z_i$ cancel when 
\begin{equation}
    \Delta= \Delta_{q,\text{opt}} = \frac1{2}\sum\limits_{\substack{i,j > \text{NN}}} V_{ij} = \frac{V_0}{2}\sum_{i,j >\text{NN}} \frac{1}{(d_{ij}/a)^6},
    \label{eq:optdetun}
\end{equation}

resulting in 

\begin{equation}
\left. H^{(2)} \right|_{\Delta = \Delta_{q,\text{opt}}} = \frac{\Omega}{2} \sum_i \sigma^x_i \prod_{i,j = \text{NN}} P_{j} +  \frac{1}{2} \sum_{i,j > \text{NN}} V_{ij} S^z_i S^z_j + \frac{\Omega^2}{4 V_0} \sum_i [\text{Many-body terms}].
\label{eq:swcorreted}
\end{equation}

\noindent with the [Many-body terms] described in Eq.~\ref{eqn:SW}. This Hamiltonian is qualitatively similar to that in Eq.~(\ref{eqn:SW}), but with smaller long-range interactions $S^z_i S^z_j$ instead of the native $n_i n_j$ interactions, due to the adopted choice of $\Delta_{q,\text{opt}}$. The long-range interactions are dominated by the contribution from next-nearest-neighbor (NNN) atoms (as $V_{ij} \propto 1/d_{ij}^6$), and due to the bipartite nature of the lattices studied here, the NNN of the $i^{\text{th}}$ atom belong to the same sublattice as the $i^{\text{th}}$ atom and thus have the same population evolution in time. For these reasons, the mean-field contribution from long-range interactions of the form $\sum_{i,j > \text{NN}} V_{ij} S^z_i S^z_j$ is roughly 1/4 the mean-field contribution of $\sum_{i,j > \text{NN}} V_{ij} n_i n_j$, and thereby reduces the deviation from the pure PXP Hamiltonian.

We emphasize that calculating the optimal value $\Delta_{q,\text{opt}}$ according to Eq.~(\ref{eq:optdetun}) requires only knowledge of $V_0$ and $d_{ij}/a$. For example, the sum in Eq.~\ref{eq:optdetun} gives $\Delta_{q,\text{opt}}/V_0 \approx 0.153,\, 0.33,\, 0.0173$ for a honeycomb lattice, a square lattice, and a one-dimensional chain respectively. For lattices where different sublattice sites are not equivalent, e.g.~Lieb and decorated honeycomb lattices, we calculate $\Delta_{q,\text{opt}}$ for both sublattices and take the average.

\begin{figure}
\includegraphics[width=\columnwidth]{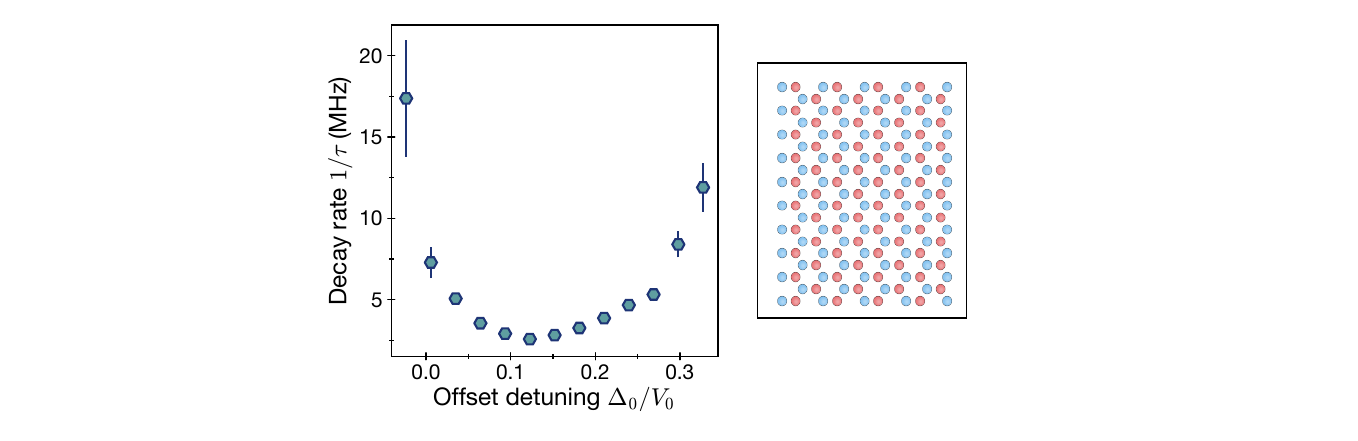}
\caption{\textbf{Optimal fixed detuning during a fixed-detuning quench.} Quenching from antiferromagnetic state $\ket{\text{AF}_1}$ to various fixed detunings $\Delta_0$ on a 162-atom honeycomb lattice with $V_0 / 2\pi = 17.1$ MHz and $\Omega/2\pi = 4.3$ MHz. The optimal fixed detuning on the honeycomb lattice is calculated to be $\Delta_{q,\text{opt}} = 1/2 \sum_{i,j > \text{NN}} V_{ij} \approx 0.153 ~V_0$. An optimum is experimentally observed here close to $\Delta_0 \approx 0.13~ V_0$, consistent with expectations from Eq.~(\ref{eq:optdetun}).
}
\label{optdetuning}
\end{figure}

In Fig.~\ref{optdetuning} we plot experimental measurements of scar decay rate $1/\tau$ under quenches to different fixed detunings $\Delta_0$ on a 162-atom honeycomb lattice. We find that the smallest decay rate is achieved at $\Delta \approx 0.13~ V_0$, close to the value of  $\Delta_{q,\text{opt}} \approx 0.153 ~V_0$ for the honeycomb lattice calculated from Eq.~(\ref{eq:optdetun}).

\subsection{2.3. Independent measurement of decay mechanisms \label{Sec:DecayRate}}

In this section we explain the expression used to describe scar decay mechanisms, and then independently corroborate the phenomenological parameters $\alpha$ and $\beta$ from the plane fit using different experimental measurements.

In the main text we used the following phenomenological expression to describe the decay rate of collective oscillations:
\begin{equation}
    \frac{1}{\tau} = \alpha\left(\frac{1}{2\pi}\sum_\text{NN} \frac{\Omega^2}{4 V_0}\right) + \beta\left(\frac{1}{2\pi}\sum_\text{NNN} V_{ij}\right) + \frac{1}{\tau_0},
    \label{eqn:appendix_thermalization}
\end{equation}
where $\alpha$, $\beta$, and $\tau_0$ are determined from the fit to the data. Physically this expression encodes the interplay of two different mechanisms that govern the behavior of $1/\tau$ and can be understood from the effective Hamiltonian~(\ref{eqn:SW}) derived in Sec.~\ref{sec:SW}. The leading term in the effective Hamiltonian~(\ref{eqn:SW}), the PXP model, leads to long-lived oscillations with significantly longer decay time  than observed for the full Rydberg Hamiltonian, both in 1D~\cite{Turner2018, Ho2019} and 2D~\cite{Michailidis2020}. 
After fixing the detuning to $\Delta_{q,\text{opt}}$ we arrive at the effective Hamiltonian in Eq.~(\ref{eq:swcorreted}), describing the PXP model perturbed by the presence of (a) hopping processes of Rydberg excitations via virtual processes that involve violation of Rydberg blockade, thus being suppressed as $\Omega^2/V_0$ at large $V_0$ and (b) longer-range interactions that scale as $V_0$, dominated by next-nearest-neighbors (NNN). Assuming that these two terms act as independent decay mechanisms, one expects two separate contributions to the decay rate that are functions of $\Omega^2/V_0$ and $V_0$ respectively, reflected by the phenomenological expression~(\ref{eqn:appendix_thermalization}).

\begin{figure}
\includegraphics[width=\columnwidth]{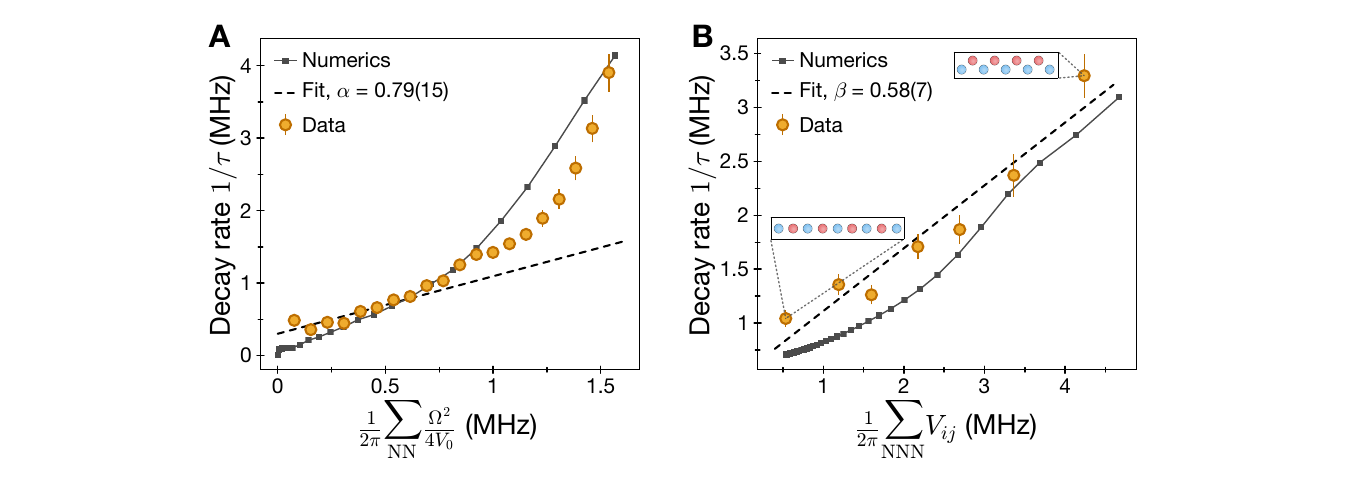}
\caption{\textbf{Independent measurement of decay parameters $\alpha$ and $\beta$.} (\textbf{A}) Measured decay rate as a function of coupling to blockade-violating states $\sim\Omega^2/(4 V_0)$, obtained by measuring at different Rabi frequencies $\Omega$ during the quench on one-dimensional 9-atom chains with a fixed $V_0/(2\pi) = 5.9\,$MHz. The linear fit (dashed line) is performed on the first 8 points, which correspond to $\Omega/V_0 < 0.5$. (\textbf{B}) Measured decay rate as a function of next-to-nearest-neighbor interactions. We prepare 9-atom chains with a variable staggering angle between neighboring sites, keeping the nearest-neighbor interaction constant at $V_0/(2\pi) = 17.1\ $MHz (insets). All error bars are given by fit uncertainties. The values for $\alpha$ and $\beta$ are consistent with the fit in the main text Fig. 2 within error bars.}
\label{alpha_beta_fits}
\end{figure}

In order to independently measure the coefficient $\alpha$, we measure the scar lifetime for different values of Rabi frequency $\Omega$, while keeping $V_0$ fixed in a 9-atom chain, thereby only changing the $\Omega^2/(4 V_0)$ term. We observe a linear dependence up to the point where $\Omega/V_0\approx0.5$, beyond which we see a strong increase of the decay rate, as the Rydberg blockade breaks down and higher-order perturbations in $\Omega/V_0$ become significant. To independently determine the value of $\beta$, we measure the scar lifetime for zigzag-shaped chains of atoms, keeping the NN spacing constant while changing the NNN spacing (Fig.~\ref{alpha_beta_fits}B), thereby only changing the NNN interaction term.

The two independent procedures described above result in values $\alpha=0.79(15)$ and $\beta=0.58(7)$, which are consistent with the values extracted from the two-dimensional fit in the main text Fig.~2 ($\alpha = 0.72(12),~ \beta = 0.58(5)$). We also perform numerical simulations of the quenches in Fig.~\ref{alpha_beta_fits} to corroborate our observations and explore imperfections of our phenomenological model. Numerical simulations of the decay rate (plotted in Fig.~\ref{alpha_beta_fits}A) agree well with the experimental data in the intermediate range of $\Omega$. However, the fine-grained theoretical curve in Fig.~\ref{alpha_beta_fits}B reveals a significant curvature for low NNN interactions, deviating from the naive linear prediction and suggesting that the phenomenological expression~(\ref{eqn:appendix_thermalization}) is an oversimplification and that the effective $\beta$ can depend on the probed range of interaction strength. We further speculate that these oversimplifications could be more dramatic in two-dimensional lattices, where e.g. the square lattice only has a small range of $V_0$ which balances the contributions from imperfect blockade and NNN interactions. Future work could explore deviations from Eq.~\ref{eqn:appendix_thermalization} and perhaps devise clever ways to suppress these decay channels.

For Fig.~2 in the main text, we also include data on lattices (Lieb, decorated honeycomb, edge-imbalanced decorated honeycomb) whose different sublattices have different imperfect blockade and NNN corrections. In these geometries, for the $x$- and $y$-axis values on the plane fit, we calculate which sublattice has the faster decay rate as given by Eq.~\ref{eqn:appendix_thermalization}, and use those values of NN imperfect blockade and NNN interactions as the $x$ and $y$ values in the plot.

\section{3. Experimental data on enhancement of scars by periodic driving\label{Sec:driven}}

\subsection{3.1. Definition of subharmonic weight\label{Sec:sub}}

In this section we describe the Fourier transform and normalization procedures for calculating $S(\omega)$. We use the in-phase component of the Fourier transform, and because the sublattice population imbalance $\mathcal{I}(t) =  \langle n\rangle_A - \langle n \rangle_B$ oscillates about a small, finite offset, we subtract the time-averaged imbalance $\overline{\mathcal{I}}$, giving

\begin{equation}
    \tilde{S}(\omega) = \frac{2}{T} \intop_0^T \text{d}t\, \left[ \mathcal{I}(t) - \overline{\mathcal{I}} \right] \cos(\omega t), 
\end{equation}

\noindent where $T$ is the longest measured quench time. Akin to the definition in \cite{Choi2017}, we then normalize by the total integrated intensity, giving

\begin{equation}
    \left|S(\omega)\right|^2 = \frac{|\tilde{S}(\omega)|^2}{2\int_0^\infty |\tilde{S}(\omega')|^2 ~\text{d}\omega'~ (T/2\pi)}.
\end{equation}

Finally, since we take a Fourier transform over a finite window $T$, to ensure the subharmonic weight is consistently defined and properly normalized, we then calculate $\left| S(\omega) \right|^2$ for a perfect subharmonic response $\mathcal{I}(t) = \cos\left[(\omega_m/2) t\right]$ and normalize such that $\left| S(\omega_m/2) \right|^2 = 1$ for this perfect subharmonic response. These normalizations yield the $\left| S(\omega) \right|^2$ that we plot throughout this work. In this way, the subharmonic weight $\left|S(\omega_m/2)\right|^2$ has a maximum of 1 which is achieved for a perfect cosine response in-phase with the drive. The intensity of the complex Fourier transform yields the same qualitative result but is broader by a factor of $\approx 2$ in the frequency domain due to the finite width of time window $T$ used in Fourier transformation.

\subsection{3.2. Robustness of subharmonic response as a function of system size}

In this section we describe the behavior of the subharmonic response as a function of the system size. A key signature of time-crystalline behavior is that the subharmonic response becomes more rigid as the system size increases \cite{Else2016,VonKeyserlingk2016}. 
 
\begin{figure*}
\includegraphics[width=\columnwidth]{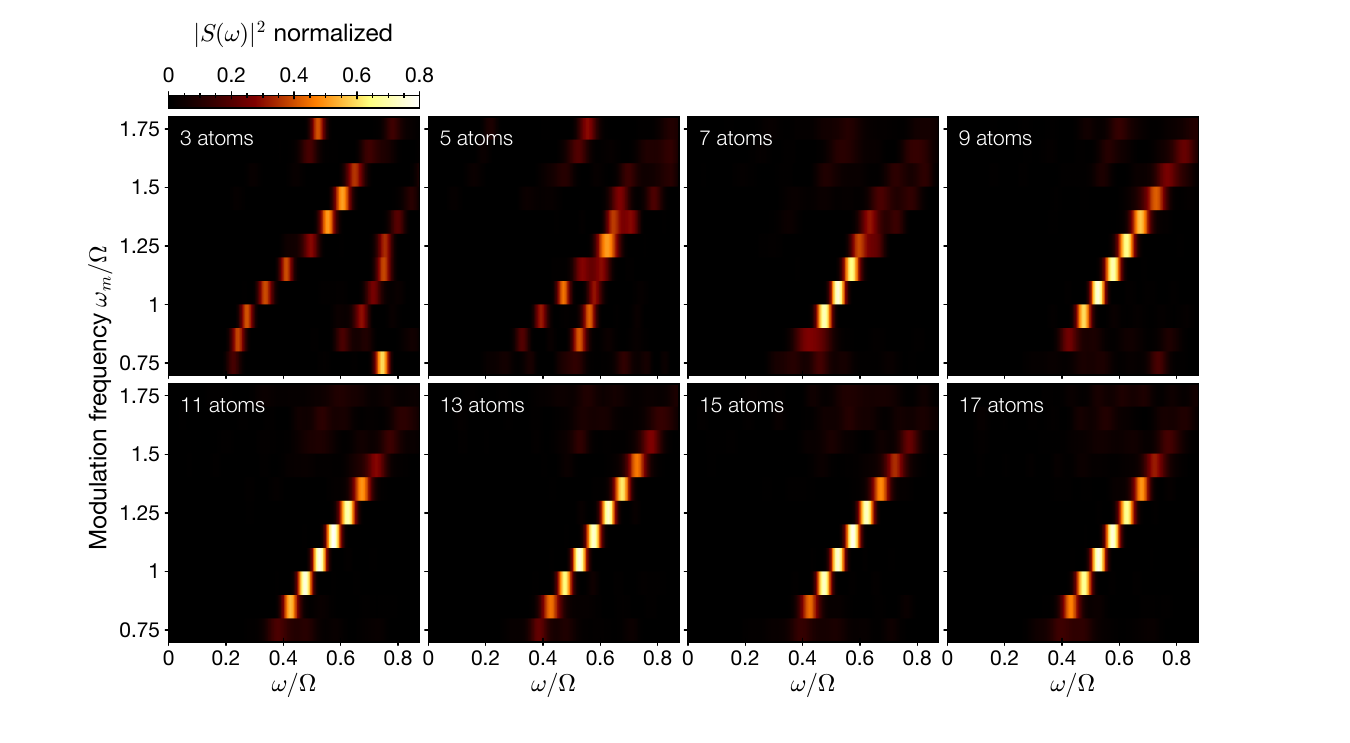}
\caption{\textbf{System-size dependence of the subharmonic response.} Fourier transform intensity $|S(\omega)|^2$ of $\braket{n}_A - \braket{n}_B$ traces for a chain of varying system size. A prominent subharmonic feature emerges and becomes more robust as the number of atoms in the chain increases, signifying that the subharmonic response is a many-body effect. All data here is a chain with $V_0 / 2\pi$ = 51 MHz, and with drive parameters $\Delta_m = \Delta_0 = 0.55 ~\Omega$.
}
\label{drivengallery}
\end{figure*}

Figure~\ref{drivengallery} plots $\left|S(\omega)\right|^2$ as a function of modulation frequency $\omega_m$ for one-dimensional chains of 3 -  17 atoms. For the 3-atom chain, a discernible subharmonic response is not observed. For the 5-atom chain, a subharmonic response is observed with $\omega_m \approx 2\times$ the natural oscillation frequency, but at larger or smaller $\omega_m$ the response splits into two separate peaks. For the 7-atom chain, the subharmonic response persists over a wider region of $\omega_m$ and with larger peak amplitude, but at sufficiently large or small $\omega_m$ the response again splits into two peaks. Finally, for chains with 9 atoms and beyond, a stable subharmonic response is observed, with large response amplitude and no discernible splitting of the central peak. 

To summarize these results quantitatively, in main text Fig.~4D we plot the subharmonic rigidity, which evaluates the robustness of the subharmonic response over a range of modulation frequencies and is defined as $\sum_{\omega_m} |S_{\omega_m}(\omega=\omega_m/2)|^2$ for $\omega_m = 0.75, 0.85, ..., 1.75 ~\Omega$. We attribute the small decrease in rigidity for the larger chains to a reduction in fidelity of the state preparation into one of the classical $\ket{\text{AF}}$ orderings. In addition to the chain data presented here, in the main text we also plot the measured subharmonic rigidity for a honeycomb lattice with sizes ranging from 9 to 200 atoms.

\begin{figure}
\includegraphics[width=\columnwidth]{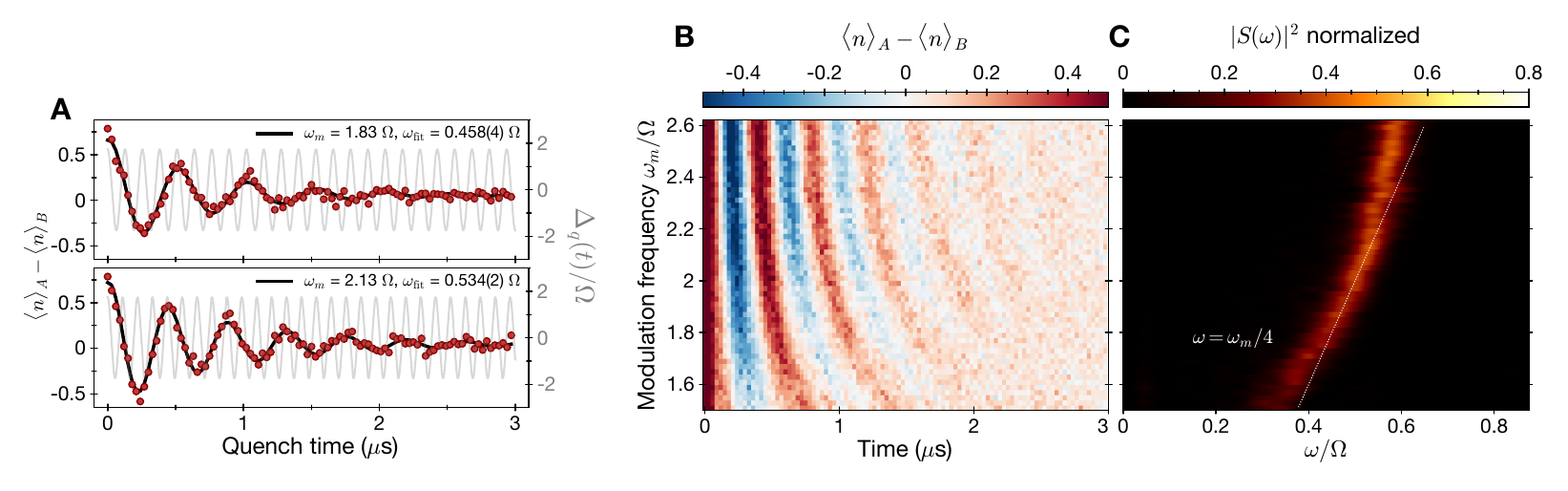}
\caption{\textbf{Signatures of a 4$^{\text{th}}$ subharmonic response.} ({\bf A}) $\braket{n}_A - \braket{n}_B$ in the presence of two different drives with modulation frequencies of $\omega_m = 1.83 ~\Omega$ and $2.13 ~ \Omega$, resulting in responses at a 4$^{\text{th}}$ subharmonic of $\omega_{\text{fit}} = 0.458(4) ~\Omega$ and $0.534(2) ~\Omega$, respectively. Data is on a 9-atom chain with $V_0 / 2\pi = 32$ MHz and drive parameters $\Delta_m = 1.75 ~\Omega$ and $\Delta_0 = 0$, which is a different drive parameter regime than those used in investigating a $2^{\text{nd}}$ subharmonic response. ({\bf B})~$\braket{n}_A - \braket{n}_B$ data for modulation frequencies from $1.51 ~\Omega$ to $2.61 ~ \Omega$ with same parameters as {\bf A}. ({\bf C}) Fourier transform intensity $|S(\omega)|^2$ of data in {\bf B}, showing signatures of a 4$^{\text{th}}$ subharmonic response (dotted white line) while seemingly not as robust as the $2^{\text{nd}}$ subharmonic response focused on in this work.
}
\label{4thsubharmonic}
\end{figure}

\subsection{3.3. Signatures of a 4$^{\text{th}}$ subharmonic response\label{sec:4th}     }

In this section we report signatures of a 4$^{\text{th}}$ subharmonic response. Figure~\ref{4thsubharmonic}A plots $\braket{n}_A - \braket{n}_B$ in the presence of two different drives with modulation frequencies of $\omega_m = 1.83 ~\Omega$ and $2.13 ~ \Omega$, resulting in responses at a 4$^{\text{th}}$ subharmonic of $\omega_{\text{fit}} = 0.458(4) ~\Omega$ and $0.534(2) ~\Omega$, respectively. Here, the quantum state synchronously returns to itself every four drive periods of $\Delta_q(t)$, as seen by comparing $\braket{n}_A - \braket{n}_B$ with the $\Delta_q(t)$ profile (gray curve).

In Fig~\ref{4thsubharmonic}B we then explore this 4$^{\text{th}}$ subharmonic response by plotting the time dynamics $\braket{n}_A - \braket{n}_B$ for modulation frequencies from $1.51 ~\Omega$ to $2.61 ~ \Omega$, and in Fig~\ref{4thsubharmonic}C plot its associated Fourier transform intensity $|S(\omega)|^2$. In panel~C we observe signatures of a 4$^{\text{th}}$ subharmonic response persisting from modulation frequencies $\omega_m$ of approximately $1.8~ \Omega$ to 2.2 $\Omega$, seemingly less robust than the $2^{\text{nd}}$ subharmonic response that is the main focus of this work. A stronger 4$^{\text{th}}$ subharmonic response may exist in other drive parameter regimes or lattice configurations (we did not search widely).

\subsection{3.4. Dependence of relaxation rate and subharmonic response on the initial state\label{sec:statedep}}

In this section we demonstrate the strong dependence of the quantum dynamics on the choice of initial state, for quenches to both fixed detunings and time-dependent detunings. Such markedly different behavior and thermalization time for different initial states can be viewed as a key signature of quantum scarring. 

\begin{figure}
\includegraphics[width=\columnwidth]{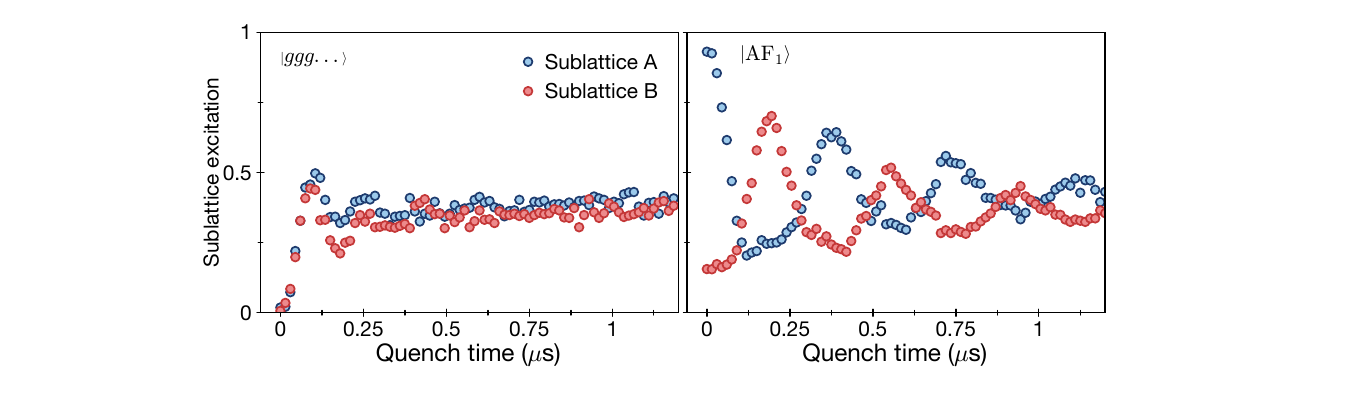}
\caption{\textbf{Initial-state dependence on dynamics.} Plotted here are fixed detuning quenches in a two-dimensional lattice (54-atom decorated honeycomb with $V_0 / 2\pi$ = 9.1 MHz and $\Omega/2\pi = 4.2$ MHz). With a $\ket{ggg...}$ initial state (left) the sublattice populations quickly equilibrate. With an $\ket{\text{AF}_1}$ initial state (right) the sublattice populations oscillate and equilibrate at a significantly slower rate, whose rate is dominated by imperfect blockade and NNN interactions as explored in the main text. 
}
\label{gggdec}
\end{figure}

\begin{figure}
\includegraphics[width=\columnwidth]{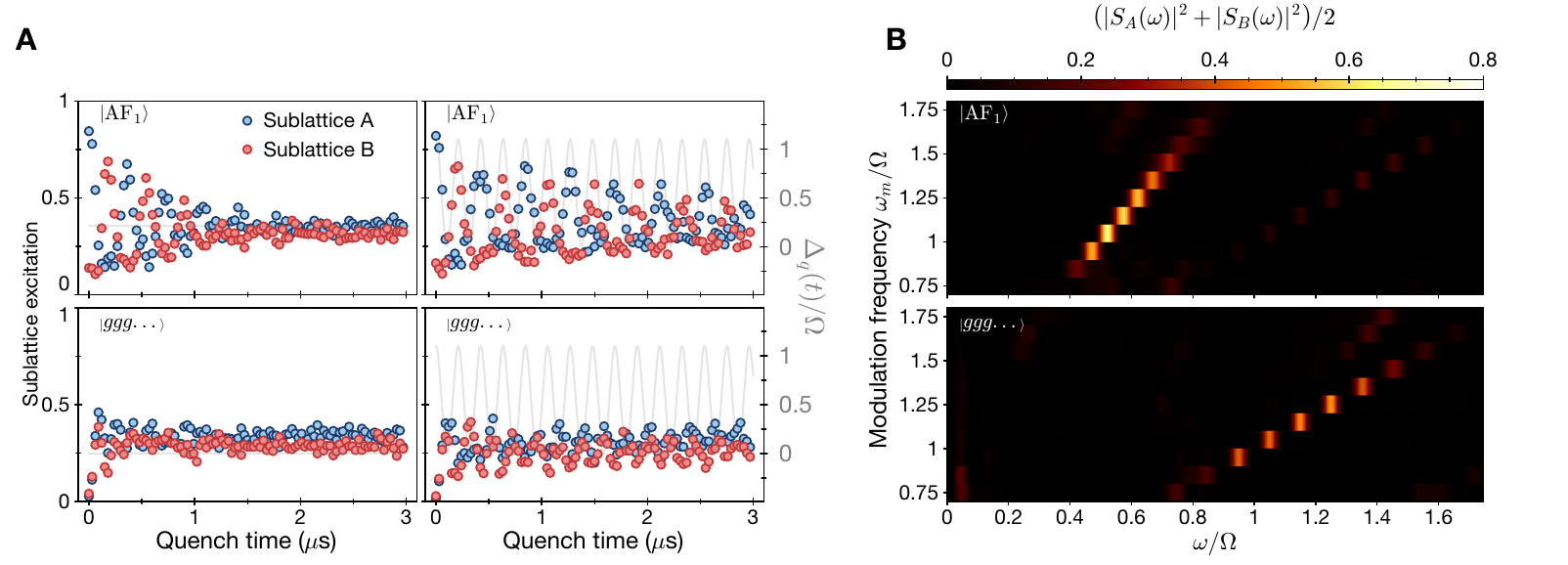}
\caption{\textbf{Response to drive for different initial states.} ({\bf A}) (Left panels) Fixed detuning quench for $\ket{\text{AF}}$ (top) and $\ket{ggg\ldots}$ (bottom) initial states, showing an initial-state dependence to the ensuing dynamics and equilibration time. (Right panels) Time-dependent quench for $\ket{\text{AF}}$ (top) and $\ket{ggg\ldots}$ (bottom) initial states. The $\ket{\text{AF}}$ state scars are prolonged and the individual sublattice response is synchronously locked to half the drive frequency, whereas the sublattice populations of the $\ket{ggg\ldots}$ state show small oscillations at the drive frequency (harmonic response). ({\bf B}) Fourier transform intensity of the individual sublattices $|S_A(\omega)|^2$ and $|S_B(\omega)|^2$, averaged together. The $\ket{\text{AF}}$ initial state (top) shows a strong subharmonic response and also a weak harmonic response (which disappears for $|S_{A-B}(\omega)|^2$ as plotted in Figure 4B of the main text). The $\ket{ggg\ldots}$ initial state (bottom) shows a harmonic response but no detectable signatures of a subharmonic response.
}
\label{gggdrive}
\end{figure}

First we present our measurement results for quenches with a static, optimal positive detuning. We plot the sublattice populations over time for an initially prepared $\ket{\text{AF}}$ state (also referred to as $\ket{\mathbb{Z}_2}$ in one dimension \cite{Turner2018}) and an initially prepared $\ket{ggg...}$ state, for a decorated honeycomb (Fig.~\ref{gggdec}) and for a one-dimensional chain (Fig.~\ref{gggdrive}). In both the two-dimensional and one-dimensional systems, the sublattice populations of the $\ket{ggg...}$ state quickly equilibrate, whereas the $\ket{\text{AF}}$ state exhibits long-lived, periodic many-body revivals. These observations experimentally confirm the initial-state dependence characteristic of quantum scarring in one and two dimensions.

In Fig.~\ref{gggdrive}A we explore the relationship between the parametric drive and quantum scarring by plotting the response of the $\ket{\text{AF}}$ and $\ket{ggg\ldots}$ states with and without a drive. For the $\ket{\text{AF}}$ state, the drive prolongs the sublattice oscillations and locks their oscillation frequency to half the drive frequency. In contrast, the sublattice populations of the $\ket{ggg\ldots}$ state still quickly equilibrate under the drive and exhibit small oscillations at the drive frequency (harmonic response). In Figure~\ref{gggdrive}B we explore these distinct responses over a range of modulation frequencies by plotting the Fourier transform intensity of the sublattice dynamics. In Fig.~4B of the main text and other figures we plot $|S(\omega)|^2 = |S_{A-B}(\omega)|^2$, but this quantity is not informative for the $\ket{ggg\ldots}$ state as it approaches zero in the thermodynamic limit. Accordingly, in Fig.~\ref{gggdrive}B we plot the average Fourier transform intensity of the individual sublattices, ($|S_A(\omega)|^2$ + $|S_B(\omega)|^2$)/2, for both the initial $\ket{\text{AF}}$ state and $\ket{ggg\ldots}$ state. We find the $\ket{\text{AF}}$ initial state exhibits a strong subharmonic response and also a weak harmonic response (which disappears for $|S_{A-B}(\omega)|^2$), whereas the $\ket{ggg\ldots}$ initial state shows a harmonic response but no detectable subharmonic response. These observations suggest that the subharmonic stabilization observed here is intertwined with the scarring behavior itself, and is distinct from conventional time crystals by this dramatic initial-state dependence even for short times / small systems.

\clearpage
\subsection{3.5. Subharmonic response with square pulse modulation\label{sec:square}}

\begin{figure}
\includegraphics[width=\columnwidth]{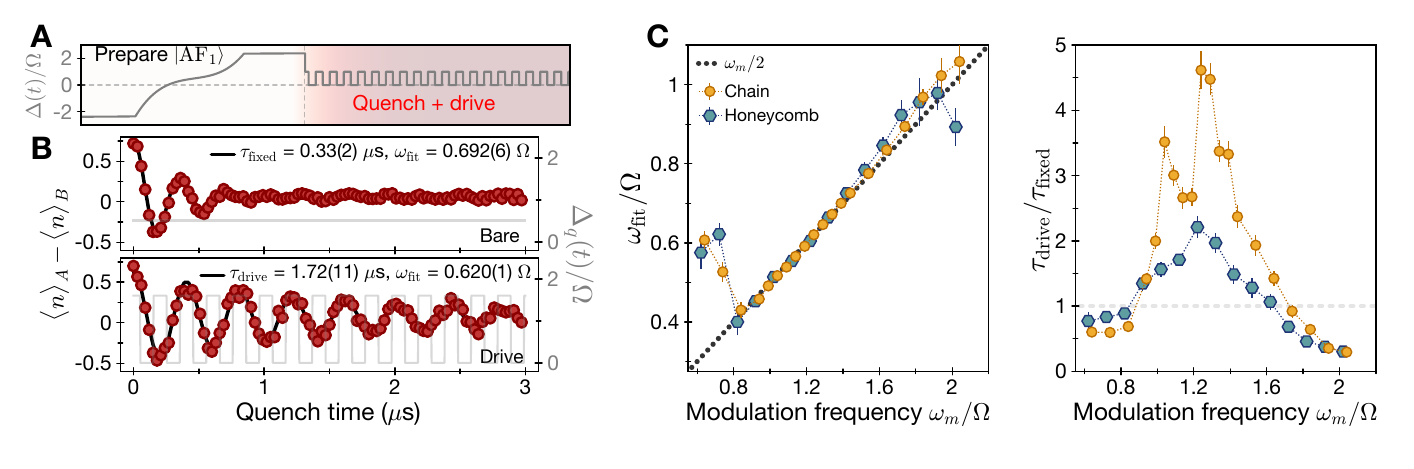}
\caption{\textbf{Subharmonic locking with square pulse modulation.} ({\bf A}) Pulse sequence with square pulse drive. ({\bf B}) Scar dynamics during a quench to a fixed optimal detuning (bare), and a time-dependent detuning (drive) with modulation frequency $\omega_m$ = 1.24 $\Omega$. The drive increases the scar lifetime and changes its frequency to $\omega_m/2$. ({\bf C}) Scar lifetime and response frequency as a function of $\omega_m$, showing a lifetime increase and subharmonic locking.
}
\label{squarepulse}
\end{figure}

In this section we demonstrate the robustness of the scar enhancement with respect to the pulse shape, here specifically for square pulses of $\Delta(t)$, as shown in Figure~\ref{squarepulse}A. Figure~\ref{squarepulse}B plots the dynamics of $\braket{n}_A - \braket{n}_B$ with a fixed detuning $\Delta_q = \Delta_{q,\text{opt}} = 0.5 ~\Omega$ (top) and a time-dependent detuning $\Delta_q(t) = \Delta_0 + \Delta_m \left(2 \Theta\left[\cos(\omega_m t)\right] - 1 \right)$ (bottom), where $\Theta$ is the Heaviside Theta Function. As with the cosine drive, the square pulse modulation increases the scar lifetime by a factor of five, from $\tau_{\text{fixed}} = 0.33(2) \,\mu$s to $\tau_{\text{drive}} = 1.72(11)\,\mu$s, and changes the oscillation frequency to be half the drive frequency of $\omega_m = 1.24\,\Omega$. In Figure~\ref{squarepulse}C we plot the fitted oscillation frequency and change in lifetime from driving as a function of the drive frequency, again finding a robust subharmonic locking to $\omega_m/2$ and accompanying lifetime increase, for a one-dimensional chain and a honeycomb lattice. Note that the chain in Fig~\ref{squarepulse}C has $V_0/2\pi = 120$ MHz, different than the $V_0/2\pi$ = 51 MHz used in Fig~3C of the main text and resulting in the different change in driven lifetime. We do not find a significant difference between the behavior of the system to cosine vs square driving, and focus on cosine driving throughout this work for consistency.

\subsection{3.6. Rationale and robustness for choice of drive parameters $\Delta_m$ and $\Delta_0$\label{sec:amp}}

In this section we discuss the choice of modulation amplitude $\Delta_m$ and offset $\Delta_0$. Largely, these values were chosen empirically, in what was observed (experimentally and numerically) to be a robust phase space.

Similar to the discussion in Section~$\ref{sec:pulse}$, preliminary hypotheses suggested that the driven stability arises in part from having extremal values of $\Delta(t)$ at times when the antiferromagnetic $\ket{\text{AF}}$ states arise, stabilizing these states as they have maximal excitation number in the blockaded subspace. Our naive hypothesis was further that we desire a cosine profile that gives $\Delta(t) \approx 0$ at times between the revival of the $\ket{\text{AF}}$ states, in order to not disrupt the scar evolution. To satisfy these conditions, we chose values of roughly $\Delta_m = \Delta_0$ and then further optimized empirically, which seemed to be close to an optimum in the various lattices and $V_0$ we measured experimentally. For the idealized PXP Hamiltonian, we find that $\omega_m = 1.33 ~\Omega$ and $\Delta_m = 2 \Delta_0 = \Omega$ appears to give oscillations which persist to hundreds of cycles. We speculate that good values for the full Rydberg Hamiltonian near $\Delta_m = \Delta_0$ instead of $\Delta_m = 2 \Delta_0$ could be a consequence of the static field of the long-range interactions, requiring a larger $\Delta_0$ to impose a static offset akin to $\Delta_{q,\text{opt}}$. We further speculate that there is an interplay between the time-dependent component of the detuning with the time-dependent component of the long-range interactions.

\begin{figure}[h]
\includegraphics[width=\columnwidth]{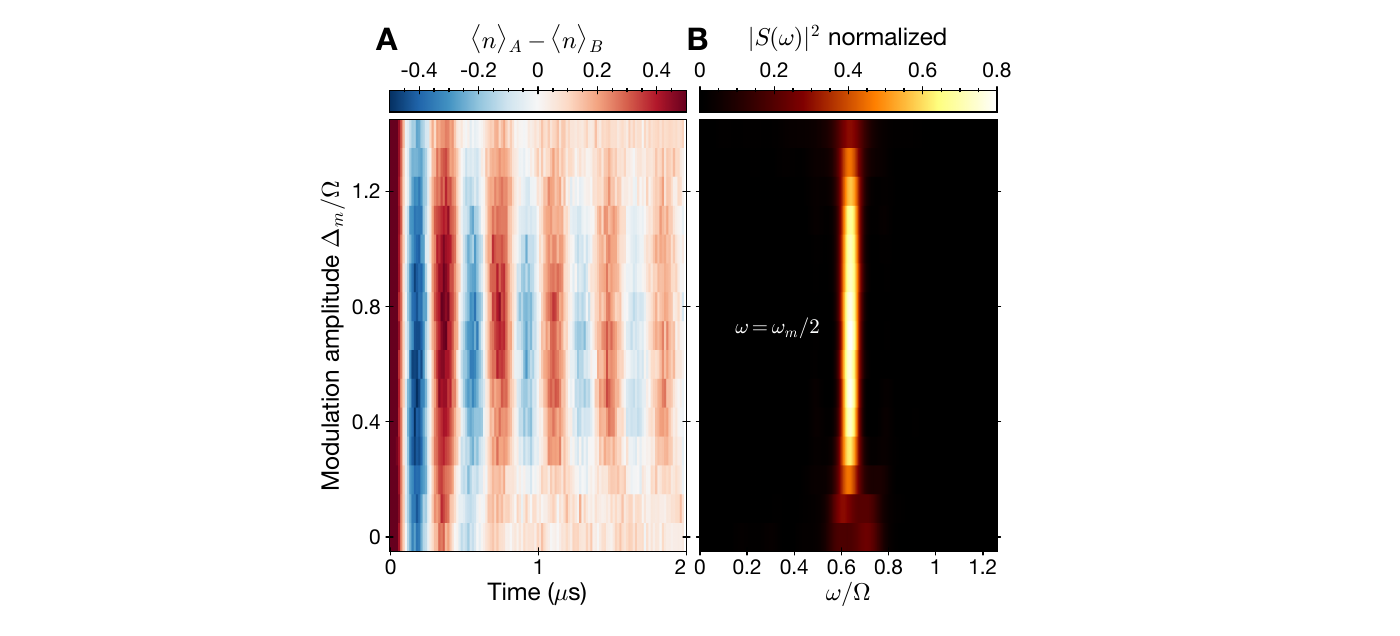}
\caption{\textbf{Subharmonic stabilization as a function of modulation amplitude $\Delta_m$}. (\textbf{A}) Dynamics of sublattice population difference after quench as a function of modulation frequency, measured on a 9-atom chain with nearest-neighbor interaction strength $V_0/2\pi = 120$ MHz = 28 $\Omega/2\pi$, detuning offset $\Delta_0$ = 0.5 $\Omega$, and modulation frequency $\omega_m = 1.28 ~\Omega$. The $\Delta_0$ we choose here is commensurate with the optimal fixed-detuning quench $\Delta_{q,\text{opt}}$ on this lattice, so the $\Delta_m = 0$ line corresponds to data for optimal undriven scars on this lattice.
(\textbf{B}) Fourier transform intensity $|S(\omega)|^2$ of data in {\bf A}. Upon applying drive amplitude $\Delta_m \approx 0.3~ \Omega$, the scar lifetime dramatically increases and exhibits a rigid subharmonic response at $\omega = \omega_m/2 = 0.64 ~\Omega$, independent of drive amplitude, before degrading at drive amplitude $\Delta_m \approx 1.3~ \Omega$.
}
\label{fig:deltam}
\end{figure}

In Fig.~\ref{fig:deltam} we plot the $\braket{n}_A - \braket{n}_B$ dynamics and associated Fourier transform for a one-dimensional chain as a function of modulation amplitude $\Delta_m$, at fixed offset $\Delta_0 = 0.5 ~\Omega$. We observe a robust subharmonic response across a wide range of $\Delta_m$, with an optimal $\Delta_m \approx 0.7 ~\Omega$.

\newpage
\section{4. Theoretical investigations of driven scars \label{Sec:theory}}

\subsection{4.1. Growth of entanglement entropy under drive\label{sec:ent}}

\begin{figure}[h]
\includegraphics[width=\columnwidth]{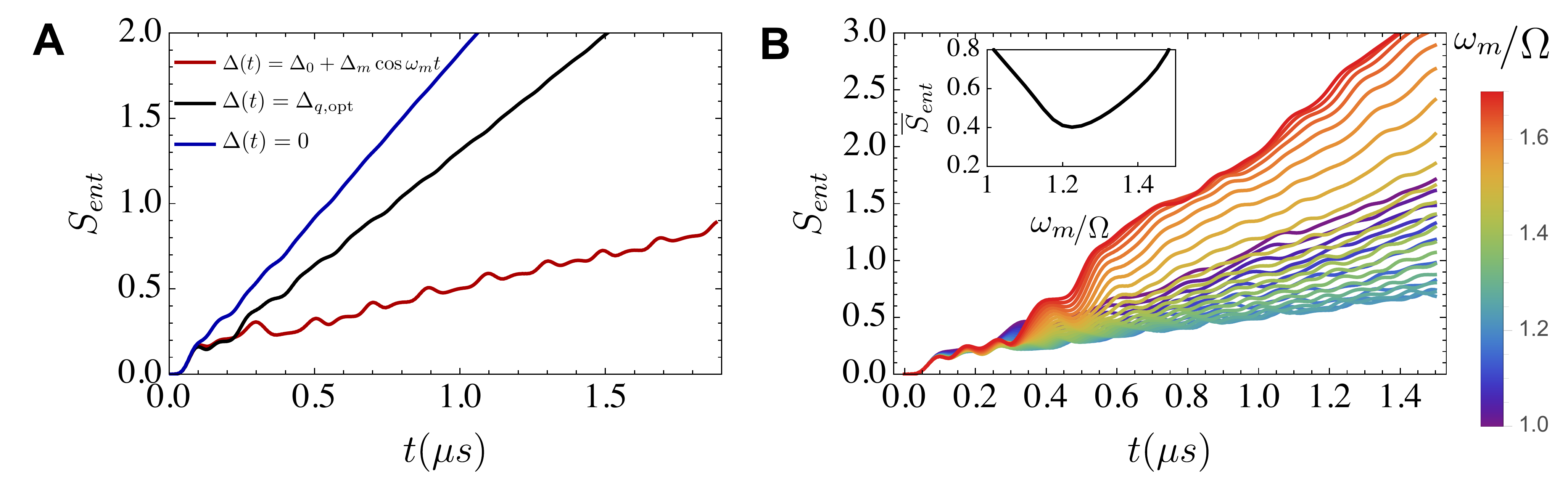}
\caption{\textbf{Entanglement entropy dynamics of the Rydberg chain for a half-chain bipartition.}  The parameters of the system are $V_{0}/2\pi = 51 $ MHz, $\Omega/2\pi= 4.2$ MHz. The time-dependent detuning amplitudes are  $\Delta_{0} = 0.55 \Omega$, $\Delta_{m} = 0.55 \Omega$.  ({\bf A}) Comparison of entanglement dynamics with harmonic detuning, optimal time-independent detuning $\Delta_{q,\text{opt}}$, and zero detuning reveals more than two-fold decrease in rate of entanglement growth due to presence of the drive. Data is shown for a 50-atom chain. The detuning parameters are $\Delta_{q,\text{opt}} = 0.0173~ V_{0} $ and $\omega_{m} = 1.2~ \Omega$.  ({\bf B}) Dependence of entanglement growth on the frequency of the drive for a 24-atom chain reveals an optimal modulation frequency that corresponds to the slowest rate of entanglement spreading.  Inset: Time averaged entropy $\overline{S}_{ent} = \frac{1}{T}\int_{0}^{T}dt\,S_{ent}$ for $T = 1.5~\mu$s shows a clear minimum around $\omega_{m}/\Omega\approx 1.225$. 
}
\label{Fig:Entanglement}
\end{figure}

The numerical data shown in Fig.~\ref{Fig:Entanglement} demonstrates the effects of the drive to the growth of bipartite entanglement entropy in the Rydberg atom chain, and therefore, provides a direct probe of the thermalization rate of the system. In Fig.~\ref{Fig:Entanglement}A we compare the growth of entropy in quenches with no detuning, optimal static detuning, and dynamical detuning, for a 50-atom chain with open boundaries. We observe that the growth of entropy in the case of the dynamical detuning is much slower compared to the cases of static and zero detuning, illustrating the qualitative difference between the driven and static systems. The large system size ensures that the entropy dynamics are not affected by finite size effects for the time period shown. The simulation is performed by applying the time-dependent variational principle on matrix product states~\cite{Haegeman2011,Haegeman2016}. The time step of the simulation is $dt = 0.002$, the truncation error is $\epsilon_{T} = 5 \cdot 10^{-9}$ and the integration is performed using a fourth-order method. The long-range interactions are truncated for distances longer than four sites. 

Figure~\ref{Fig:Entanglement}B shows the dependence of bipartite entanglement entropy growth on the frequency of the drive for a 24-atom chain with open boundaries. The calculation is performed using second-order Trotterized time evolution with time step $dt = 0.001$ applied to the full wave function. The slope of the entropy growth achieves a minimum value at $\omega_m \approx 1.225~ \Omega$, similar to the optimal $\omega_m$ observed experimentally in main text Fig~3C for the 9-atom chain.

\subsection{4.2. Stabilization of pure PXP \label{sec:pxp}}

\begin{figure}[h]
\includegraphics[width=\columnwidth]{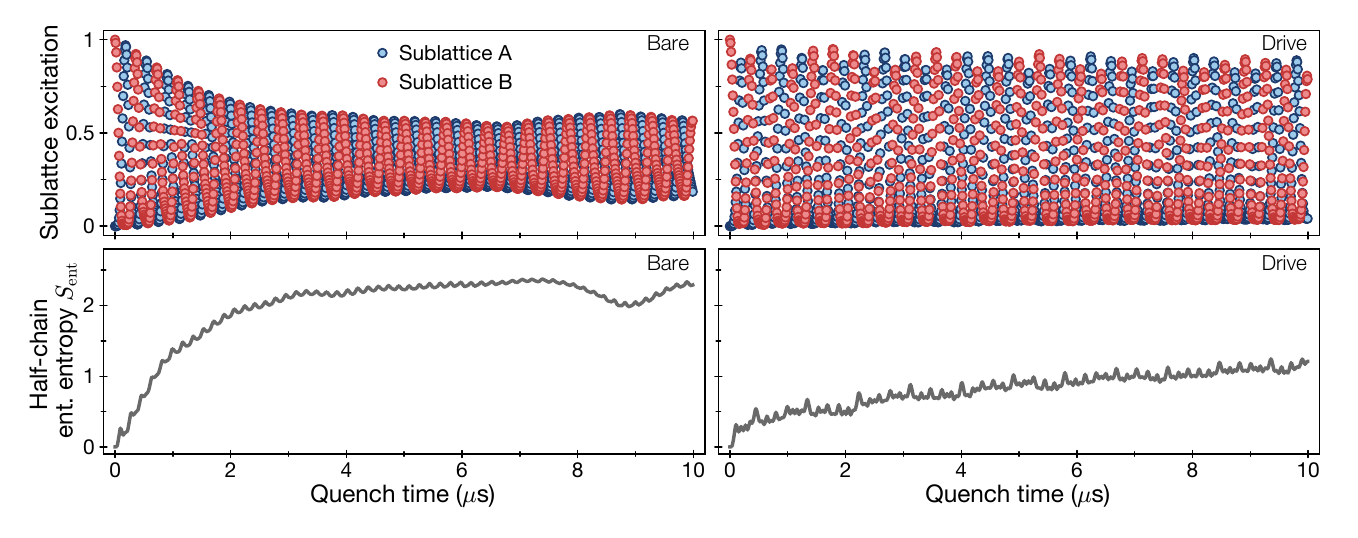}
\caption{\textbf{Stabilization of pure PXP model under drive (numerics).} (Top) Sublattice excitation probability for undriven (left) and driven (right) PXP model. (Bottom) Entanglement entropy across midway bipartition for undriven (left) and driven (right) PXP model. Numerics are calculated for a 22-atom chain with periodic boundary conditions and the timescale is set by $\Omega/2\pi = 4.2$ MHz. ``Bare'' is a conventional quench to $\Delta = 0$ and ``Drive'' is a quench to $\Delta = \Delta_0 + \Delta_m \cos\left(\omega_m t\right)$, with drive parameters $\Delta_0 = 0.5~ \Omega$, $\Delta_m = 1.0 ~ \Omega$, and $\omega_m = 1.33~ \Omega$. These plots show that the cosine drive allows to delay the onset of thermalization even for the ``idealized'' PXP model, which describes perfect nearest-neighbor blockade ($V_0 = \infty$) with no long-range interactions.
}
\label{Fig:PXPdrive}
\end{figure}

Figure~\ref{Fig:PXPdrive} demonstrates that the dynamics of the idealized PXP model, 
\begin{equation}
H_{\text{PXP}}(t) = \sum_{i}\left(\frac{\Omega}{2} P_{i-1}\sigma^{x}_{i}P_{i+1}-\Delta(t)n_{i}\right),
\end{equation}

are also stabilized by the cosine drive.
We use a Krylov method to evolve a 22-atom chain with periodic boundary conditions in the blockaded Hilbert space. Both the slow growth of bipartite entanglement entropy and the slow decay of sublattice excitation revivals provide evidence for a suppression of thermalization mechanisms in the driven system. This result also illustrates that the effect of time-dependent detuning cannot be simply attributed to the cancellation of the long-range interactions as the drive is able to further suppress thermalization of the idealized PXP model.

\newpage
\subsection{4.3. Analysis of pulsed model  \label{sec:pulse}}
Here we detail the pulsed model of scar stabilization presented in the main text, corresponding to a simplified model (we assume infinitely sharp detuning pulses and idealized PXP interactions) that qualitatively reproduces key experimental observations of extended lifetime and subharmonic locking from scar states, as well as strong initial-state dependence of the phenomenon. We note that the combined concepts of pulsed Floquet driving and Rydberg atoms has been explored theoretically, although in regimes distant from the work here \cite{Fan2020, Mukherjee2020, Mukherjee2020b, Mizuta2020a}.

The pulsed model is given by the Hamiltonian
\begin{align}
    H(t) & = H_{\textrm{PXP}} + \theta N \sum_{n \in \mathbb{Z} } \delta(t-n\tau^{-}),
\end{align}
which consists of $\tau$-periodic delta-function `kicks' of the detuning $N = \sum_i n_i$  with amplitude $\theta$, on top of the PXP Hamiltonian. This can be thought of as an idealized, limiting case of the experimental driving where the detuning is applied instantaneously once per period. This time-dependent Hamiltonian generates the Floquet unitary

\begin{align}
    U_F(\theta, \tau) &= e^{-i \theta N} e^{-i \tau H_{\textrm{PXP}}},
\end{align}

which comprises of two parts: evolution under $H_{\textrm{PXP}}$ for time $\tau$, and then an application of $N$ for an angle $\theta$.
For a fine-tuned evolution time $\tau_c \approx 0.755 \times 2\pi~ \Omega^{-1}$ the first step $e^{-i \tau_c H_{\textrm{PXP}}}$ acts like an approximate spin-flip between the $\ket{\textrm{AF}_1}$ and $\ket{\textrm{AF}_2}$ product states, but otherwise generically serves to generate entanglement for initial states.

\begin{figure}
    \centering
    \includegraphics[width=\textwidth]{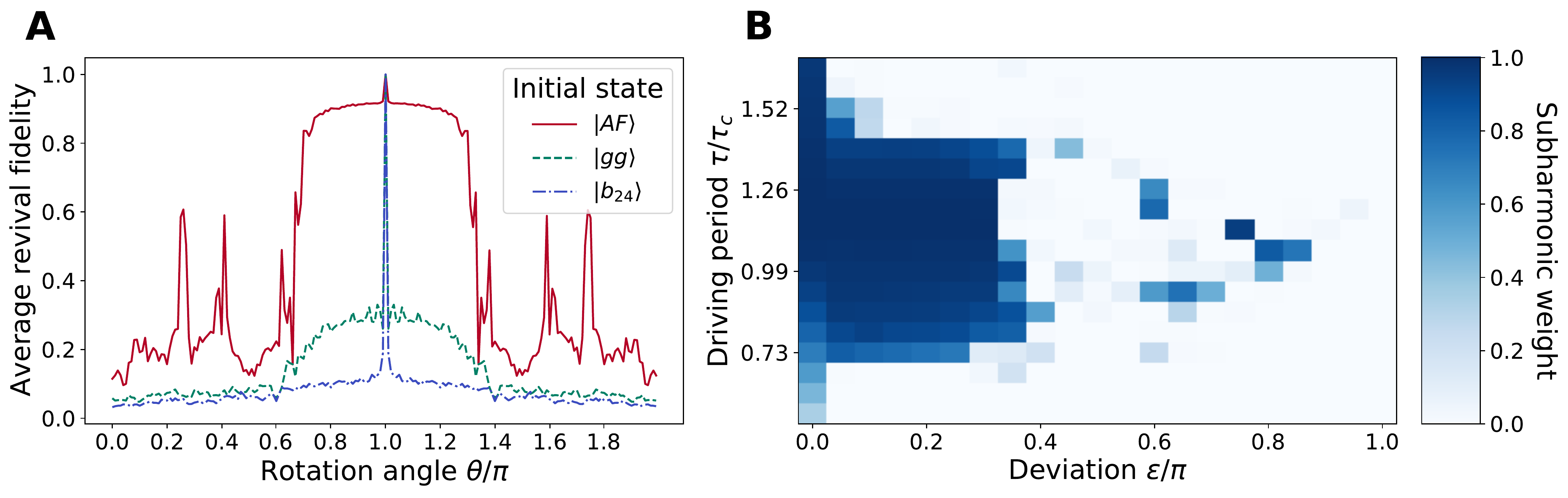}
    \caption{\textbf{State-dependent subharmonic revivals in the pulsed drive model.} \textbf{(A)} Many-body revivals under Floquet unitary $U_F$ for $\tau=\tau_c$ and varying $\varepsilon = \pi - \theta$ for a $L=14$ chain with periodic boundary conditions. Revivals were calculated by taking the average of $\vert \bra{ \text{AF}}U_F(\pi + \varepsilon,\tau)^{2n}\ket{\text{AF}} \vert^2$ for $n=1,2,...,100$. \textbf{(B)} Here, we depict the dependence of subharmonic weight on the rotation angle $\tau$ under $H_{\textrm{PXP}}$ and the deviation $\varepsilon$ from the perfect echo point $\theta = \pi$, calculated for $N=400$ driving periods. We see oscillations persist to larger $\varepsilon$ for $\tau$ near $\tau_c$.}
    \label{fig:diff_initstates}
\end{figure}

The Floquet unitary, parameterized by $(\theta,\tau)$, harbors a special point $\theta=\pi$. There the drive reverses dynamics generated by $H_{\textrm{PXP}}$ perfectly after two driving periods. Specifically, the PXP Hamiltonian has a particle-hole symmetry under $e^{-i\pi N}$, i.e. $e^{-i\pi N} H_{\textrm{PXP}} e^{i\pi N}=-H_{\textrm{PXP}}$ (because $\sigma^z_i \sigma^x_i \sigma^z_i = - \sigma^x_i$). As such, the application of PXP during the first driving period is exactly undone during the second driving period, i.e.~$U_F^2 = \mathbb{I}$. This is essentially a many-body echo, and produces perfect subharmonic revivals for all initial states for any value of $\tau$. However, we find that away from the $\theta=\pi$ point where such an echo is no longer perfect, the $\ket{\textrm{AF}_1}$ and $\ket{\textrm{AF}_2}$ states nevertheless still exhibit substantial many-body revivals for a wide range of deviations $\varepsilon = \theta - \pi$, at fixed $\tau =  \tau_c$. Indeed as can be seen in (Fig. \ref{fig:diff_initstates}), there is a plateau of stability for $\theta$ near $\pi$ for which the oscillations from the $\ket{\textrm{AF}_1}$ states persist beyond 100s of Floquet periods.

The pulsed model also displays subharmonic locking, notably for the $|\text{AF}\rangle$ initial states but not others like $|ggg\cdots\rangle$ (Fig.~\ref{fig:diff_initstates}B). To probe the dependence on $\tau$ and $\varepsilon$, we compute the weight of  subharmonic response in the power spectrum of $\langle n(t) \rangle_A - \langle n(t) \rangle_B$,
defined above in the main text and references \cite{Choi2017,Zhang2017}. 
Numerical results show that robust oscillations for $|\text{AF}\rangle$ persist until very late times (100-1000s of Floquet periods), for a wide range of $\tau$ near $\tau_c$.

To explain the origin of this wide window of stability, we rewrite the Floquet unitary as
\begin{align}
    U_F(\theta,\tau) &= e^{-i\varepsilon N} X = e^{-i \varepsilon \sum_{\langle i j \rangle} \sigma^z_i \sigma^z_j} X, \nonumber \\
    X &\equiv e^{-i \pi N} e^{-i \tau H_{\textrm{PXP}}}.
\end{align}
Here, we make two important conceptual observations. First, the operator $X$ is the Floquet unitary at the special point $\theta=\pi$, and as such it squares to one, i.e.~$X^2 = \mathbb{I}$. Second, we notice that since we operate within the blockaded subspace, $\sum_{\langle i j \rangle} n_i n_j = 0$, we can rewrite $\sum_i n_i = \sum_{\langle i j \rangle} \sigma^z_i \sigma^z_j + \mathrm{const.}$, justifying the second equality up to an irrelevant global phase.
This Floquet unitary is of the form studied in the context of discrete time crystals (DTC) where conventionally $X$ is a global spin-flip $\prod_i \sigma_i^x$  \cite{Else2016,Khemani2016, Else2020}. Importantly, however, $X$ in our case is not a product of simple on-site operators but instead generates entangled dynamics.

However, $X$'s action implements an approximate global spin flip between the product states $|\text{AF}_1\rangle$ and $|\text{AF}_2\rangle$ when $\tau = \tau_c$, as a result of the special quantum scarring properties that $H_\text{PXP}$ possesses. Furthermore, $N$ serves to stabilize these states, as they are contained within the two dimensional blockaded ground state manifold  of $\varepsilon \sum_{\langle i j \rangle} \sigma^z_i \sigma^z_j$ which is separated from the rest of the spectrum by a constant gap $\varepsilon$. Thus, loosely speaking, these two product states simply oscillate between one another (at stroboscopic times). The robustness of the subharmonic response across a wide parameter range is likely a result of the gap, which protects the  oscillations against additional generic small perturbations to the drive (as long as they still respect the time-translation symmetry, i.e.~the drive is still Floquet in nature) \cite{VonKeyserlingk2016, Else2017}. Note that such analysis does not carry over to other initial product states, and so we do not expect robust many-body revivals from them.

\begin{figure}[b]
    \includegraphics[width=0.8\textwidth]{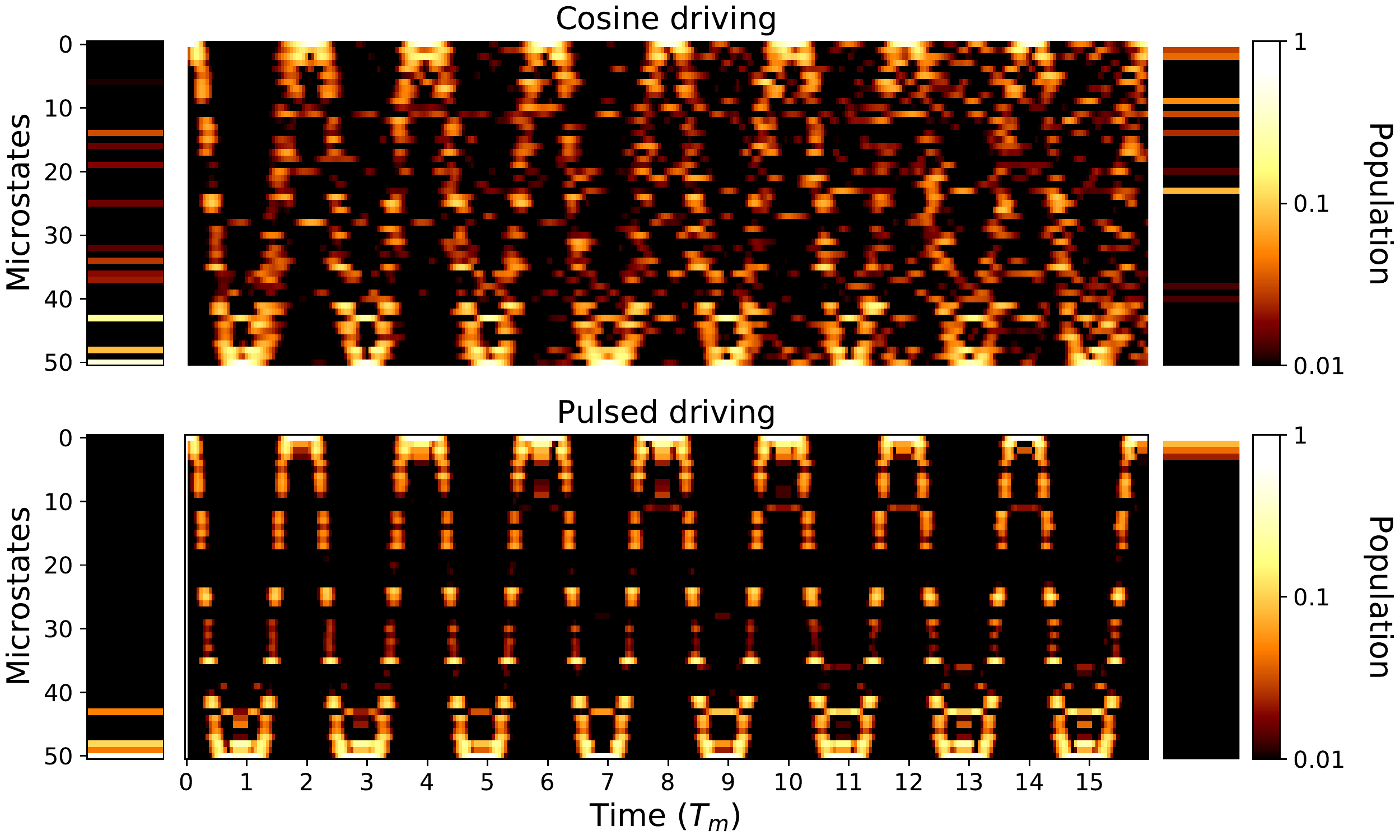}
    \caption{\textbf{Numerical simulations of microstate population dynamics}.  We plot the microstate distribution for (top) cosine driving ($\Delta_0=\Delta_m=0.55\Omega$) and (bottom) pulsed driving ($\varepsilon=0.5$) at driving frequency $\omega_m=1.15\Omega$ for the 1D $L=9$ chain. The states are ordered by their Hamming distance from the $|\text{AF}_1\rangle$ state. The right and left columns depict a decomposition into microstates of the two symmetric and anti-symmetric superpositions of the two Floquet eigenstates with largest overlap with $|\textrm{AF}_1\rangle, |\textrm{AF}_2\rangle$ states respectively. Dynamics (center column) appear to be largely explained by these two eigenstates, as can be seen from the agreement between microstate populations at stroboscopic times. We note that the microstates populated at stroboscopic times, for both the simulations involving cosine and pulsed driving, are in qualitative agreement with those observed for the experimental protocol, as shown in Fig.~3D of the main text.
    }
    \label{fig:microstate_comparisions}
\end{figure}

The pulsed model also provides an avenue by which to understand the microstate plot in Fig.~3D in the main text, which focuses on a 1D chain as we similarly do so below. The plot shows that driving induces stable oscillations between two states which have large populations in the antiferromagnetic states, but also acquires a signficant amplitude in other microstates. Empirically, we observe that these additional microstates tend to have large values of $N$, and are hence microstates that have smallest energy difference from the $\ket{\text{AF}}$ states as measured by $\varepsilon \sum_{\langle ij \rangle} \sigma^z_i \sigma^z_j$. The pulsed model also predicts this behavior (Fig.~\ref{fig:microstate_comparisions}).

The interesting behavior of the pulsed model presented above warrants future, more detailed theoretical analysis. We emphasize however that many open questions remain, including: the role of significant next-nearest-neighbor interactions, the observed frequency range of locking (main text Fig.~4B), the multi-peak structure seen in the driven lifetime of the edge-imbalanced decorated honeycomb (main text Fig.~3C), and the 4$^{th}$ subharmonic response (Section~\ref{sec:4th}).
Furthermore, although the pulsed model reproduces key phenomenological aspects, the precise connection between the pulsed driving and continuous driving implemented experimentally is left for future work.

\clearpage
\section{5. Tabulation of system and drive parameters used in main text \label{sec:tabulation}}

\begin{table}[h]
\centering
\begin{tabular}{|l|l|l|l|}
 \hline
 Figure & Lattice& Geometry parameters & Quench / drive parameters \\
 \hline
  & & & $\Omega/2\pi$ = 4.2 MHz \\
 \hline
 Fig 1B,C & Honeycomb  & 85 atoms, $V_0/2\pi = 9.1$ MHz  & $\Delta_0 = \Delta_{q, \text{opt}} = 0.15 ~V_0$ \\
 Fig 2 & Chain  & 9 atoms, $V_0$ = varied  & $\Delta_0 = \Delta_{q, \text{opt}} = 0.017~ V_0$ \\
 Fig 2 & Square  & 49 atoms, $V_0$ = varied  & $\Delta_0 = \Delta_{q, \text{opt}} = 0.33 ~V_0$ \\
 Fig 2 & Honeycomb  & 85 atoms, $V_0$ = varied  & $\Delta_0 = \Delta_{q, \text{opt}} = 0.15~ V_0$ \\
 Fig 2 & Lieb  & 129 atoms, $V_0/2\pi = 9.1$ MHz  & $\Delta_0 = \Delta_{q, \text{opt}} = 0.20 ~V_0$ \\
 Fig 2 & Dec. hon.$^a$  & 54 atoms, $V_0$ = varied  & $\Delta_0 = \Delta_{q, \text{opt}} = 0.10 ~V_0$ \\
 Fig 2 & EIDH$^b$  & 66 atoms, $V_0$ = varied  & $\Delta_0 = \Delta_{q, \text{opt}} = 0.10 ~V_0$ \\
 Fig 3B & Chain bare  & 9 atoms, $V_0/2\pi = 120$ MHz & $\Delta_0 = \Delta_{q, \text{opt}} = 0.50 ~\Omega$\\
 Fig 3B & Chain drive & Same as bare & $\omega_m = 1.24 ~\Omega$, $\Delta_0 = 0.85~ \Omega$, $\Delta_m = 0.98 ~\Omega$ \\
 Fig 3C & Chain   & 9 atoms, $V_0/2\pi = 51$ MHz & $\omega_m = $ varied, $\Delta_0 = 0.55~ \Omega$, $\Delta_m = 0.55 ~\Omega$ \\
 Fig 3C & Honeycomb   & 41 atoms, $V_0/2\pi = 24$ MHz & $\omega_m = $ varied, $\Delta_0 = 0.87~ \Omega$, $\Delta_m = 0.87 ~\Omega$ \\
 Fig 3C & EIDH$^b$   & 66 atoms, $V_0/2\pi = 29$ MHz & $\omega_m = $ varied, $\Delta_0 = 0.78~ \Omega$, $\Delta_m = 0.98 ~\Omega$ \\
 Fig 3D & Chain bare  & 9 atoms, $V_0/2\pi = 51$ MHz & $\Delta_0 = \Delta_{q, \text{opt}} = 0.21 ~\Omega$\\
 Fig 3D & Chain drive & Same as bare & $\omega_m = 1.15~ \Omega$, $\Delta_0 = 0.55~ \Omega$, $\Delta_m = 0.55 ~\Omega$ \\
 Fig 3E & Chain bare  & 16 atoms, $V_0/2\pi = 51$ MHz & $\Delta_0 = \Delta_{q, \text{opt}} = 0.21 ~\Omega$\\
 Fig 3E & Chain drive & Same as bare & $\omega_m = 1.20~ \Omega$, $\Delta_0 = 0.55~ \Omega$, $\Delta_m = 0.55 ~\Omega$ \\
 Fig 4A,B & Chain   & 9 atoms, $V_0/2\pi = 51$ MHz & $\omega_m =$ varied, $\Delta_0 = 0.55~ \Omega$, $\Delta_m = 0.55 ~\Omega$ \\
 Fig 4C & Chain   & 9 atoms, $V_0/2\pi = $ varied & $\omega_m =$ varied, $\Delta_0 = 0.55~ \Omega$, $\Delta_m = 0.55 ~\Omega$ \\
 Fig 4C & Honeycomb   & 41 atoms, $V_0/2\pi = $ varied & $\omega_m =$ varied, $\Delta_0 = 0.87~ \Omega$, $\Delta_m = 0.87 ~\Omega$ \\
 Fig 4D & Chain  & 3 - 17 atoms, $V_0/2\pi = 51$ MHz & $\omega_m =$ varied, $\Delta_0 = 0.55~ \Omega$, $\Delta_m = 0.55 ~\Omega$ \\
 Fig 4D & Honeycomb  & 9 - 200 atoms, $V_0/2\pi = 17$ MHz & $\omega_m =$ varied, $\Delta_0 = 0.87~ \Omega$, $\Delta_m = 0.87 ~\Omega$ \\

 \hline
\end{tabular}
\caption{\textbf{Tabulation of system and drive parameters used in the main text}. $^a$Dec. hon. stands for decorated honeycomb. $^b$EIDH stands for edge-imbalanced decorated honeycomb. Varied indicates that this parameter is varied in the plot.}
\label{parametertable}
\end{table}

\clearpage
\section{6. Tabulation of 51-dimensional Hilbert space from main text Figure~3D \label{sec:tabulation2}}
\begin{table}[h]
\centering
\begin{tabular}{|c|c|c|c|}
 \hline
 Index & Microstate & Index & Microstate \\
 \hline
1 & 1 0 1 0 1 0 1 0 1 & 27 & 0 0 0 0 0 0 1 0 0 \\
2 & 1 0 0 0 1 0 1 0 1 & 28 & 1 0 1 0 0 1 0 1 0 \\
3 & 1 0 1 0 1 0 1 0 0 & 29 & 1 0 0 1 0 1 0 0 1 \\
4 & 1 0 1 0 0 0 1 0 1 & 30 & 0 0 0 0 1 0 0 1 0 \\
5 & 0 0 1 0 1 0 0 0 1 & 31 & 0 1 0 0 0 0 0 0 1 \\
6 & 1 0 0 0 0 0 1 0 1 & 32 & 1 0 0 1 0 0 0 0 0 \\
7 & 0 0 1 0 1 0 1 0 0 & 33 & 0 0 1 0 0 0 0 1 0 \\
8 & 1 0 0 0 1 0 0 0 1 & 34 & 1 0 0 0 0 1 0 0 0 \\
9 & 1 0 1 0 0 0 1 0 0 & 35 & 0 0 1 0 0 1 0 0 0 \\
10 & 0 0 0 0 1 0 1 0 1 & 36 & 0 0 0 0 0 0 0 0 0 \\
11 & 0 1 0 0 1 0 1 0 1 & 37 & 0 0 1 0 0 1 0 1 0 \\
12 & 1 0 1 0 0 1 0 0 1 & 38 & 0 1 0 1 0 0 0 0 1 \\
13 & 1 0 0 0 0 0 0 0 1 & 39 & 1 0 0 1 0 1 0 0 0 \\
14 & 1 0 0 0 0 0 1 0 0 & 40 & 1 0 0 1 0 0 0 1 0 \\
15 & 1 0 0 0 1 0 0 0 0 & 41 & 0 1 0 0 1 0 0 1 0 \\
16 & 0 0 1 0 0 0 1 0 0 & 42 & 0 0 0 0 0 0 0 1 0 \\
17 & 0 0 0 0 0 0 1 0 1 & 43 & 0 0 0 1 0 0 0 0 0 \\
18 & 0 0 1 0 1 0 0 0 0 & 44 & 0 1 0 1 0 1 0 0 1 \\
19 & 0 0 0 1 0 0 1 0 1 & 45 & 0 0 0 0 0 1 0 1 0 \\
20 & 0 1 0 0 1 0 0 0 1 & 46 & 0 1 0 0 0 0 0 1 0 \\
21 & 0 1 0 0 0 0 1 0 1 & 47 & 0 1 0 0 0 1 0 0 0 \\
22 & 0 1 0 0 1 0 1 0 0 & 48 & 0 0 0 1 0 1 0 0 0 \\
23 & 0 0 1 0 0 1 0 0 1 & 49 & 0 1 0 0 0 1 0 1 0 \\
24 & 1 0 0 0 0 1 0 0 1 & 50 & 0 1 0 1 0 1 0 0 0 \\
25 & 0 0 0 0 1 0 0 0 0 & 51 & 0 1 0 1 0 1 0 1 0 \\
26 & 1 0 0 0 0 0 0 0 0 &  &  \\
    \hline
\end{tabular}
\caption{\textbf{Tabulation of microstates in main text Figure~3D}. For the 9-atom chain, the $2^9$-dimensional Hilbert space is first reduced to 89 states by discarding states that violate the Rydberg blockade constraint, giving rise to the so-called ``constrained Hilbert space''. The Hilbert space dimension is then further reduced from 89 to 51 by grouping left-right symmetric pairs of microstates. Finally, the microstates are ordered by $n_A - n_B$, or equivalently by Hamming distance from $\ket{\text{AF}_1}$, and within a given cluster of $n_A - n_B$, states are then ordered by $n_A + n_B$ (although this ordering is still not completely unique). ``0'' represents ground state $\ket{g}$ and ``1'' represents Rydberg state $\ket{r}$.}
\label{microstatetable}
\end{table}

\end{document}